\documentclass[aps,prl,twocolumn,longbibliography]{revtex4-2}
\usepackage{graphicx,comment,color,tikz,amsmath,amssymb}

\usepackage{physics}
\usepackage{xcolor}
\usepackage{ulem}

\usepackage{wasysym} 
\usepackage[colorlinks=true, citecolor=dodgerblue,linkcolor=dodgerblue,urlcolor=dodgerblue,pdftitle={CDLakes}]{hyperref}

\definecolor{orange(ryb)}{HTML}{FFA500}
\definecolor{dodgerblue}{HTML}{1E90FF}
\usetikzlibrary{snakes}

\def\starZs{\: \tikz[baseline={([yshift=-.5ex]current bounding box.center)},scale=1.3]{
\draw (0,0) -- (0.25,0) node {$Z$} -- (0.75,0) node {$Z$} -- (1,0);
\draw (0.5,-0.5) -- (0.5,-0.25) node {$Z$} -- (0.5,0.25) node {$Z$} -- (0.5,0.5); \:}
}

\def\starYs{\: \tikz[baseline={([yshift=-.5ex]current bounding box.center)},scale=1.3]{
\draw (0,0) -- (0.25,0) node {$Y$} -- (0.75,0) node {$Y$} -- (1,0);
\draw (0.5,-0.5) -- (0.5,-0.25) node {$Y$} -- (0.5,0.25) node {$Y$} -- (0.5,0.5); \:}
}

\def\starYZs{\: \tikz[baseline={([yshift=-.5ex]current bounding box.center)},scale=1.3]{
\draw (0,0) -- (0.25,0) node {$Z$} -- (0.75,0) node {$Z$} -- (1,0);
\draw (0.5,-0.5) -- (0.5,-0.25) node {$Z$} -- (0.5,0.25) node {$Y$} -- (0.5,0.5); \:}
}

\def\starXZs{\: \tikz[baseline={([yshift=-.5ex]current bounding box.center)},scale=1.3]{
\draw (0,0) -- (0.25,0) node {$Z$} -- (0.75,0) node {$Z$} -- (1,0);
\draw (0.5,-0.5) -- (0.5,-0.25) node {$Z$} -- (0.5,0.25) node {$X$} -- (0.5,0.5); \:}
}

\def\starIXs{\: \tikz[baseline={([yshift=-.5ex]current bounding box.center)},scale=1.3]{
\draw (0,0) -- (0.25,0) node {$X$} -- (0.75,0) node {$X$} -- (1,0);
\draw (0.5,-0.5) -- (0.5,-0.25) node {$X$} -- (0.5,0.5); \:}
}

\def\starYYYZ{\: \tikz[baseline={([yshift=-.5ex]current bounding box.center)},scale=1.3]{
\draw (0,0) -- (0.25,0) node {$Y$} -- (0.75,0) node {$Y$} -- (1,0);
\draw (0.5,-0.5) -- (0.5,-0.25) node {$Z$} -- (0.5,0.25) node {$Y$} -- (0.5,0.5); \:}
}

\def\starYYZs{\: \tikz[baseline={([yshift=-.5ex]current bounding box.center)},scale=1.3]{
\draw (0,0) -- (0.25,0) node {$Z$} -- (0.75,0) node {$Y$} -- (1,0);
\draw (0.5,-0.5) -- (0.5,-0.25) node {$Z$} -- (0.5,0.25) node {$Y$} -- (0.5,0.5); \:}
}

\def\starXstarZs{\: \tikz[baseline={([yshift=-.5ex]current bounding box.center)},scale=1.3]{
\draw (0,0) -- (0.25,0) node {$Z$} -- (0.75,0) node {$X$} -- (1.25,0) node {$Z$} -- (1.5,0);
\draw (0.5,-0.5) -- (0.5,-0.25) node {$Z$} -- (0.5,0.25) node {$Z$} -- (0.5,0.5); 
\draw (1,-0.5) -- (1,-0.25) node {$Z$} -- (1,0.25) node {$Z$} -- (1,0.5); 
\:}
}

\def\starXstarXstarZs{\: \tikz[baseline={([yshift=-.5ex]current bounding box.center)},scale=1.3]{
\draw (0,0) -- (0.25,0) node {$Z$} -- (0.75,0) node {$X$} -- (1.25,0) node {$X$} -- (1.75,0) node {$Z$} -- (2,0);
\draw (0.5,-0.5) -- (0.5,-0.25) node {$Z$} -- (0.5,0.25) node {$Z$} -- (0.5,0.5); 
\draw (1,-0.5) -- (1,-0.25) node {$Z$} -- (1,0.25) node {$Z$} -- (1,0.5); 
\draw (1.5,-0.5) -- (1.5,-0.25) node {$Z$} -- (1.5,0.25) node {$Z$} -- (1.5,0.5); 
\:}
}

\def\starXstarYZs{\: \tikz[baseline={([yshift=-.5ex]current bounding box.center)},scale=1.3]{
\draw (0,0) -- (0.25,0) node {$Z$} -- (0.75,0) node {$X$} -- (1.25,0) node {$Z$} -- (1.5,0);
\draw (0.5,-0.5) -- (0.5,-0.25) node {$Z$} -- (0.5,0.25) node {$Y$} -- (0.5,0.5); 
\draw (1,-0.5) -- (1,-0.25) node {$Z$} -- (1,0.25) node {$Z$} -- (1,0.5); 
\:}
}

\def\starXstarYYZs{\: \tikz[baseline={([yshift=-.5ex]current bounding box.center)},scale=1.3]{
\draw (0,0) -- (0.25,0) node {$Y$} -- (0.75,0) node {$X$} -- (1.25,0) node {$Z$} -- (1.5,0);
\draw (0.5,-0.5) -- (0.5,-0.25) node {$Z$} -- (0.5,0.25) node {$Y$} -- (0.5,0.5); 
\draw (1,-0.5) -- (1,-0.25) node {$Z$} -- (1,0.25) node {$Z$} -- (1,0.5); 
\:}
}

\def\loopXs{\: \tikz[baseline={([yshift=-.5ex]current bounding box.center)},scale=1.75]{
\draw (0,0) -- (0.25,0) node {$X$} -- (0.5,0) -- (0.5,0.25) node {$X$} -- (0.5,0.5) -- (0.25,0.5) node {$X$} -- (0,0.5) -- (0,0.25) node {$X$} -- (0,0);
\:}
}

\begin{document}
\title{Shortcuts to Analog Preparation of Non-Equilibrium Quantum Lakes}
\author{Nik O. Gjonbalaj}
\email{nikgjonbalaj@g.harvard.edu}
\author{Rahul Sahay}
\author{Susanne F. Yelin}
\affiliation{Department of Physics, Harvard University, Cambridge, Massachusetts 02138, USA}

\date{\today}

\begin{abstract}
    The dynamical preparation of exotic many-body quantum states is a persistent goal of analog quantum simulation, often limited by experimental coherence times.
    Recently, it was shown that fast, non-adiabatic Hamiltonian parameter sweeps can create finite-size ``lakes'' of quantum order in certain settings, independent of what is present in the ground state phase diagram.
    Here, we show that going further out of equilibrium via external driving can substantially accelerate the preparation of these quantum lakes.
    Concretely, when lakes can be prepared, \textit{existing} counterdiabatic driving techniques---originally designed to target the ground state---instead \textit{naturally} target the lakes state.
    We demonstrate this both for an illustrative single qutrit
    and a model of a $\mathbb{Z}_2$ Rydberg quantum spin liquid.
    In the latter case, we construct experimental drive sequences that accelerate preparation by almost an order of magnitude at fixed laser power.
    We conclude by using a Landau-Ginzburg model to provide a semi-classical picture for how our method accelerates state preparation.
\end{abstract}

\maketitle

\textbf{Introduction.} Preparing exotic quantum many-body states using Hamiltonian dynamics is a central goal of analog quantum simulation \cite{Altman_2021,Preskill_2018}.
The paradigmatic tool in this quest is adiabatic state preparation, where one sweeps a Hamiltonian parameter at a rate much smaller than the spectral gap \cite{Bachmann_2017}.
While these sweeps can be prohibitively slow, various ``shortcuts to adiabaticity'' 
have been developed to accelerate this process, 
keeping state preparation within experimental coherence times and enabling further experimentation 
\cite{Chen_2010,Torrontegui_2013,Guery_Odelin_2019,del_Campo_2019}.

\begin{figure}[!t]
	\centering	
    \includegraphics[width=\linewidth]{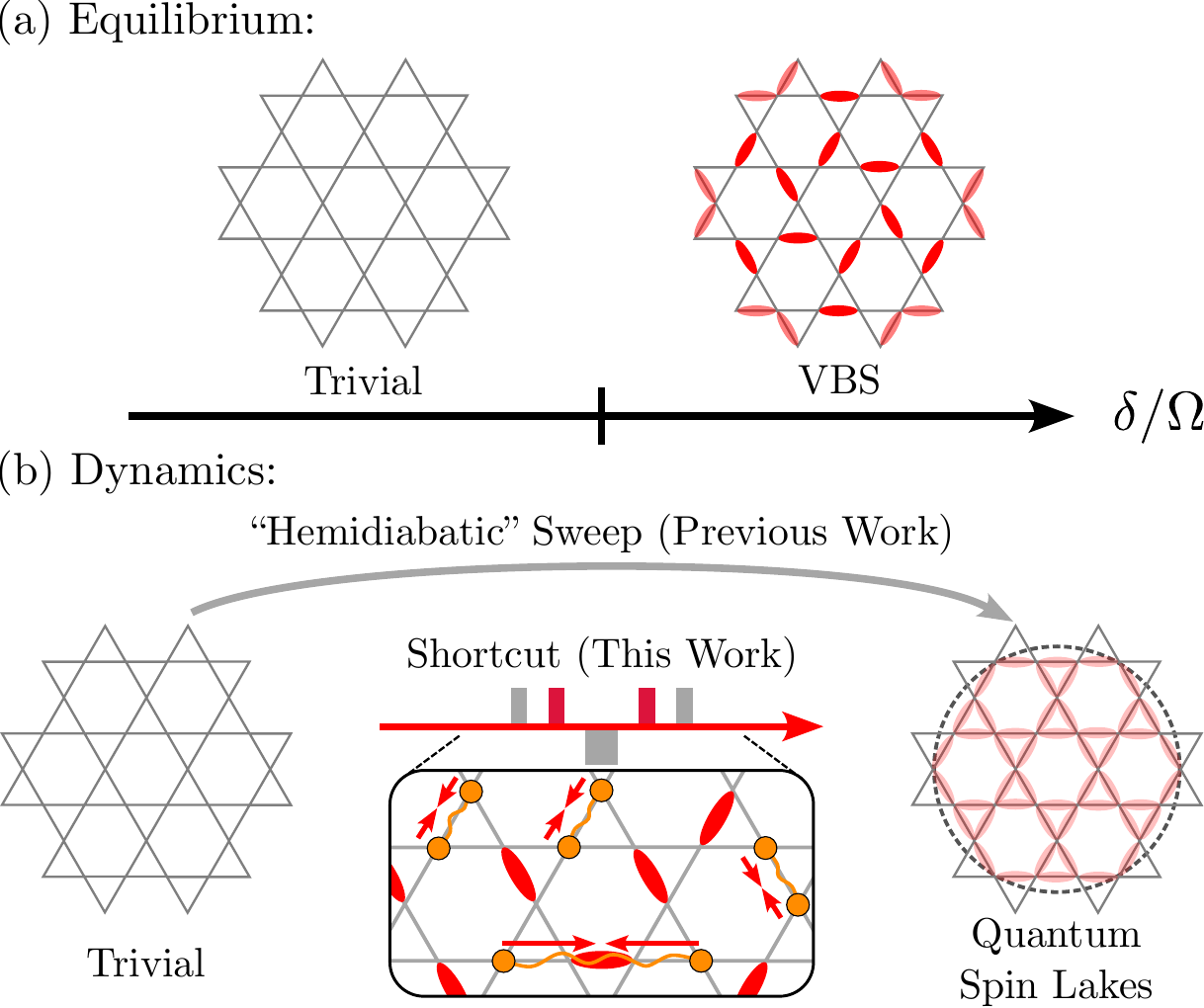}
    
    \caption{\textbf{Quantum Lakes from Driving.} (a) Experimental phase diagrams may only host non-exotic, classically described orders.
    For example, for Rydberg atoms on the links of the kagome lattice, the phase diagram as a function of detuning over laser power $\delta/\Omega$ is dominated by trivial and valence bond solid (VBS) phases (with excited atoms depicted as red dimers); exotic orders occupy small parameter regimes and can be destabilized by experimental imperfections~\cite{Verresen_2021}.
    (b) Previous work \cite{Semeghini_2021, Giudici_2022, Sahay_2023,Cheng_2023} showed that
    increasing $\delta/\Omega$ 
    at a ``hemidiabatic'' rate---between the usual sudden and adiabatic regimes---can prepare finite-size ``lakes'' of exotic (e.g. spin liquid) order independent of ground state order.
    Here, we construct CD driving protocols that efficiently force ``defects'' [e.g. vacancies in the dimer covering (orange)] out of the initial state, accelerating lakes preparation by almost an order of magnitude.
    }
	\label{fig:figure 0}
\end{figure}

Nevertheless, adiabatic sweeps and their 
shortcuts have their limitations.
Indeed, they generally require experimental models to host the target state as a ground state.
Such regimes are not guaranteed to exist and, when they do, are often sensitive to experimental
imperfections 
[Fig.~\ref{fig:figure 0}(a)].
However, a recent Rydberg atom experiment \cite{Semeghini_2021} and related theoretical works \cite{Verresen_2021,Giudici_2022,Sahay_2023,Cheng_2023} highlighted that, in some cases, non-equilibrium parameter sweeps can prepare finite-size ``lakes'' of quantum order 
not present in the ground state phase diagram.
This phenomenon is a consequence of a ``hemidiabatic'' dynamical regime 
that sits 
between the more traditional adiabatic and sudden (quench) regimes [Fig.~\ref{fig:figure 0}(b)] \cite{Sahay_2023,Sahay_youtube}.
Crucially, realizing this regime only requires engineering broad energy scales of the Hamiltonian,
not the exact ground state order.
Given limited experimental coherence times, taking full advantage of this preparation scheme demands the development of ``shortcuts to hemidiabaticity''; otherwise, state preparation may consume the entire run time or not be possible at all.

\begin{figure*}[!t]
	\centering
	
	\includegraphics[width=0.9\linewidth]{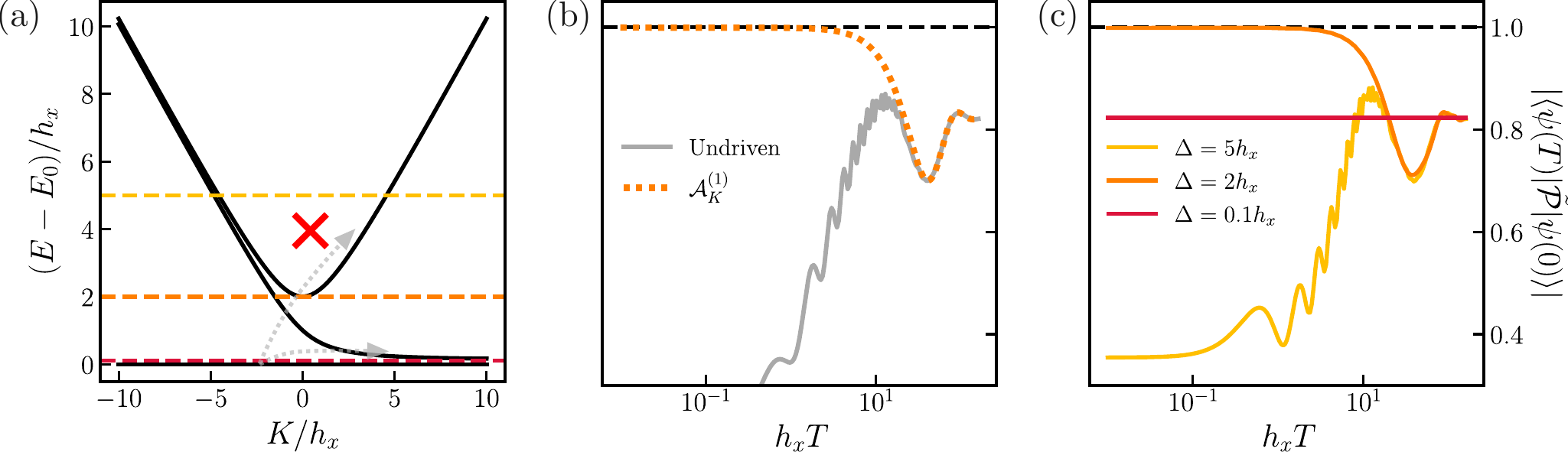}

	\caption{\textbf{Targeting Hemidiabaticity in a Qutrit.} 
    (a) We consider a qutrit as the simplest example of hemidiabaticity and show its energy spectrum (black) as a function of $K$ with $h_z = h_x/15$ [Eq.~\eqref{eq:qutrit}].
    Ref.~\cite{Sahay_2023} showed that by increasing $K$ at an \textit{intermediate} hemidiabatic rate, 
    one remains adiabatic relative to the second state (transitions are suppressed) and sudden relative to the first (represented with dotted gray arrows).
    (b) In this work, we show this effect can be reproduced and accelerated by CD driving.
    We first reproduce results of \cite{Sahay_2023} as a benchmark by simulating an undriven sweep in gray.
    We plot the overlap of the evolved state with the initial state after being projected into the low-energy subspace (spanned by $\ket{\pm1}$) as a function of total sweep time $T$, showing a peak for hemidiabatic sweep rates ($h_x T \approx 10$).
    We build on these results by driving with the approximate AGP $\mathcal{A}_K^{(1)}$ (dotted orange) which both amplifies the peak and extends it to faster sweeps.
    (c) This effect can be understood by driving with gapped AGPs, which \textit{exactly} cancel all transitions above energy $\Delta$ and allow all transitions below it.
    We plot all choices of $\Delta$ as dashed colored lines in (a).
    For large $\Delta$ (yellow), the driving only slightly changes the dynamics, while small $\Delta$ (red) reproduces adiabatic dynamics.
    In contrast, intermediate $\Delta$ (orange) reproduces the approximate CD driving.
    }
	\label{fig:qutrit}
\end{figure*}

In this work, we show that existing shortcuts to adiabaticity can \textit{naturally} target the hemidiabatic regime.
In particular, we demonstrate that standard approximations to counterdiabatic (CD) driving~\cite{Demirplak_2003,Demirplak_2005,Berry_2009,Sels_2017,Claeys_2019,Kolodrubetz_2017,Ljubotina_2022,Gjonbalaj_2022,Villazon_2021,Cepaite_2023,Morawetz_2024,hartmann2019rapid,hegade2022portfolio,barone2024counterdiabatic,Chandarana_2022,Lawrence_2025,Schindler_2024,Wurtz_2022,del_Campo_2013} prepare the non-equilibrium lakes state instead of the ground state.
We first theoretically verify this claim by 
simulating a simple qutrit and a spin liquid model from a recent Rydberg atom experiment \cite{Verresen_2021,Semeghini_2021}.
Turning to the problem of practical experimental implementation, we demonstrate that CD-inspired pulse sequences 
can speed up state preparation by nearly an order of magnitude in this model.
Finally, we discuss how this picture generalizes to models beyond spin liquids using an effective Landau-Ginzburg theory.

\textbf{CD Driving in a Qutrit.}
First, we discuss the simplest model exhibiting the hemidiabatic regime---the single qutrit from Ref.~\cite{Sahay_2023}---and show that existing approximate CD driving techniques naturally target the state prepared by a hemidiabatic sweep.
The model Hamiltonian can be expressed as
\begin{align}
    H = - K \mathcal{Z}^2 - h_x \mathcal{X} - h_z \mathcal{Z} ,
    \label{eq:qutrit}
\end{align}
where $\mathcal{X},\mathcal{Y},\mathcal{Z}$ are the spin-1 Pauli matrices.
We choose $h_z > 0$ and $h_z/h_x \ll 1$ and label $\mathcal{Z}$ eigenstates as $\ket{0},\ket{\pm1}$.

\textit{Undriven Case.}
We start by reviewing the model's undriven dynamics, studied in Ref.~\cite{Sahay_2023}.
Note that when $K$ is large and negative (positive), the ground state of the model is $\approx \ket{0}$ ($\ket{1}$).
As such, a truly adiabatic sweep from $K = -\infty$ to $+\infty$ will naturally prepare $\ket{1}$.
However, by examining the system's energy levels 
[black lines in Fig.~\ref{fig:qutrit}(a)], it is clear that an adiabatic sweep will be difficult due to the small gap between the ground state and the first excited state
for $K > 0$.
Instead, notice that there is a regime of quantum dynamics where one sweeps at a rate 
adiabatic relative to the splitting between the ground state and the second excited state 
but sudden with respect to that of the ground state and the first.
This 
is 
the \textit{hemidiabatic regime}.

Ref.~\cite{Sahay_2023} showed that an ansatz for the state created after such a sweep is given by the projection $\mathcal{P}$ of the initial state into the low-energy subspace for large positive $K$, spanned by $\ket{\pm 1}$. 
Since any initial $K \ll 0$ implies that $\ket{\psi(0)} \approx \ket{0} + \varepsilon (\ket{1} + \ket{-1})$ where $\varepsilon \sim |h_x/K|$, the prepared state is approximately $\mathcal{P} \ket{\psi(0)} \propto \frac{1}{\sqrt{2}} (\ket{1} + \ket{-1})$, a state inaccessible using adiabatic sweeps.
This picture is substantiated by the numerics shown in Fig.~\ref{fig:qutrit}(b). 
We linearly sweep $K$ from $-20 h_x$ to $20 h_x$ in time $T$ and plot the overlap with the (normalized) target state $\tilde{\mathcal{P}} \ket{\psi(0)}$ at the end of the sweep.
The overlap for this undriven sweep (shown in gray) is maximized for sweeps of intermediate length
while sudden and adiabatic sweeps perform strictly worse.

\textit{Driven Case.}
We now consider the system's behavior under approximate CD driving.
Recall that CD driving evolves the system under the time-dependent Hamiltonian $H_{\mathrm{CD}}(t) = H(t) + \dot{K} \mathcal{A}_{K}(t)$, where $\mathcal{A}_K$ (known as the adiabatic gauge potential, or AGP) is an external drive designed to cancel all transitions away from the adiabatic trajectory \cite{SM}.
While $\mathcal{A}_K$ in the qutrit can be implemented exactly, it generally becomes 
nonlocal in many-body systems and must 
be locally approximated.
Our goal is now to demonstrate that approximate CD driving naturally targets the hemidiabatic regime and prepares the projected state $\tilde{\mathcal{P}} \ket{\psi(0)}$.

\nocite{Cong_2024,Pandey_2020,bukov2018reinforcement,hegade2021shortcuts,morawetz2025universalcounterdiabaticdriving,finzgar2025counterdiabaticdrivingperformanceguarantees,landau1932theorie,zener1932non,stuckelberg1932theorie,majorana1932atomi,li2024quantumcounterdiabaticdrivinglocal,lukin2024quantumquenchdynamicsshortcut,tenpy,hauschild2024tensornetworkpythontenpy,Chandran_2013,Fredenhagen_1983,Fredenhagen_1986,1986_MARCU,Gregor_2011}

We start by considering a particular approximation scheme: the perturbative variational method from Refs.~\cite{Sels_2017,Claeys_2019}.
To lowest order, the method approximates the AGP as $\mathcal{A}_K^{(1)} = i \alpha [H, \partial_K H] = i h_x \alpha [\mathcal{X},\mathcal{Z}^2]$, where $\alpha$ can be determined analytically \cite{Sels_2017,SM} and the driving can be implemented experimentally via a Floquet protocol \cite{Claeys_2019,SM}.
The results of driving under this AGP are shown as the dotted line in Fig.~\ref{fig:qutrit}(b). 
Strikingly, the hemidiabatic peak in the undriven curve has been amplified and extended to arbitrarily sudden sweeps!
(Ref.~\cite{Villazon_2021} observed a similar effect).

\textit{Spectral Understanding.}
We now show that this behavior results from a certain ``gapped'' structure of the approximate AGP.
It has already been noted that the approximation above 
cannot effectively cancel transitions below some energy scale $\Delta$, only being effective for larger transitions \cite{Claeys_2019,SM}.
This structure is elucidated in Fig.~\ref{fig:qutrit}(c) by driving with an exactly gapped AGP that only cancels transitions above $\Delta$:
\begin{align}
    \bra{m} \mathcal{A}_{K}^{\Delta} \ket{n} \equiv \Theta(|E_m - E_n| - \Delta) \bra{m} \mathcal{A}_{K} \ket{n} ,
    \label{eq:exact gapped agp}
\end{align}
where $E_n$ ($\ket{n}$) are the instantaneous eigenvalues(vectors) of $H(t)$ and $\Theta$ is the Heaviside step function.
When $\Delta = 5h_x$ and the AGP cancels few transitions, the results nearly reproduce the undriven sweep.
When $\Delta = 0.1h_x$ and the AGP cancels all transitions, 
the adiabatic result is extended to all sweep rates.
Finally, when $\Delta = 2h_x$,
the driving reproduces the first order result from Fig.~\ref{fig:qutrit}(b),
supporting the intuition that hemidiabaticity 
results from suppressing large transitions while allowing small ones.
This gapped structure is a generic feature of many AGP approximations \cite{del_Campo_2012,Damski_2014,Villazon_2021,Cepaite_2023,Morawetz_2024,Wurtz_2022,Hastings_Wen_2005,Bachmann_2017} 
and is moreover a necessary feature of AGPs connecting different (topological) phases of matter which are defined not to be connected by any finite-time local dynamics \cite{Hastings_Wen_2005,Chen_Wen_2010,Bachmann_2017}.

\textbf{Quantum Many-Body System.}
We now demonstrate that similar results extend to the many-body setting.
In this context, the hemidiabatic sweep
will be between 
distinct phases of matter.
Crucially, the 
prepared state
will exhibit an order distinct from either of these phases.
We first show, as a theoretical point, that approximate CD driving systematically targets the hemidiabatic state before constructing drives that can be more easily implemented in experiments.

We concretely consider the 
PXP model of Rydberg atoms placed on the links of the kagome lattice (i.e. the sites of the ruby lattice) \cite{Semeghini_2021, Verresen_2021, Browaeys_2020}:
\begin{align}
    H = \frac{\Omega}{2} \sum_i P X_i P - \delta \sum_i n_i , 
    \label{eq:ruby PXP Hamiltonian}
    \quad
    \begin{tikzpicture}[scale = 0.9, baseline={([yshift=-.5ex]current bounding box.center)}]
    \foreach \i in {0,...,1}{
        \foreach \j in {0,...,1}{
            \draw[gray] ({\i + \j * 1/2}, {\j * 0.866025404}) -- ({\i + \j * 1/2 + 1/2}, {\j * 0.866025404}) -- ({\i + \j * 1/2 + 1/2*1/2}, {\j * 0.866025404 + 1/2*0.866025404}) -- cycle;
            \draw[gray] ({\i + \j * 1/2}, {\j * 0.866025404}) -- ({\i + \j * 1/2 - 1/2}, {\j * 0.866025404}) -- ({\i + \j * 1/2 - 1/2*1/2}, {\j * 0.866025404 - 1/2*0.866025404}) -- cycle;
            \filldraw  ({\i + \j * 1/2 + 1/4}, {\j * 0.866025404}) circle (1 pt);
            \filldraw  ({\i + \j * 1/2 - 1/4}, {\j * 0.866025404}) circle (1 pt);
            \filldraw  ({\i + \j * 1/2 + 1/8 }, {\j * 0.866025404 + 1/4 * 0.866025404}) circle (1 pt);
            \filldraw  ({\i + \j * 1/2 - 1/8 }, {\j * 0.866025404 - 1/4 * 0.866025404}) circle (1 pt);
            \filldraw  ({\i + \j * 1/2 + 1/8 + 1/4 }, {\j * 0.866025404 + 1/4 * 0.866025404}) circle (1 pt);
            \filldraw  ({\i + \j * 1/2 - 1/8 -1/4}, {\j * 0.866025404 - 1/4 * 0.866025404}) circle (1 pt);
        }
    }a
    \draw[red, fill=red, fill opacity=0.1, dashed, thick](1/4,0) circle (0.51);
    \filldraw[red] (1/4, 0) circle (1 pt);
    \draw [-stealth, thick] (1/4,0) -- (1/4 - 0.44, 0.25);
    \draw [stealth-stealth, line width = 0.1 mm] (5/4 + 0.1 + 0.04,-0.1 + 0.01 ) -- (5/4 + 0.25 + 0.025, 0.25 - 0.1 + 0.025 );
    \node at ({-1/4 -1/6}, 0.5) {\normalsize $R_b$};
    \node at (5/4 + 0.25 + 0.2, 0.25 - 0.15 ) {\normalsize $a$};
    \end{tikzpicture}
\end{align}
where $n_i = (1+Z_i)/2$ counts the number of atoms in the Rydberg state and $X_i,Y_i,Z_i$ are the spin-1/2 Pauli matrices on qubit $i$. 
Here, $P \equiv \prod_{i, j: |r_i - r_j| \leq R_b} \left( 1 - n_i n_j \right)$ projects out states where two atoms within a blockade radius $R_b$ are both in the Rydberg state.
\footnotetext[2]{
While this approximation does affect the phase diagram, hemidiabatic dynamics are largely unaffected \cite{Sahay_2023}, and one can modify the driving schemes below to include the full $(R_b/r)^6$ tails.} 
This blockade constraint arises energetically as a consequence of the $(R_b/r)^6$ van der Waals interaction between Rydberg atoms, approximated above to be infinite within $R_b$ and zero outside
\cite{Lukin_2001,Jaksch_2000,Ga_tan_2009,Urban_2009,Note2}.
It is convenient to represent atoms in the Rydberg state as dimers on a kagome link, i.e. 
$\ket{ \begin{tikzpicture}[scale = 0.5, baseline = {([yshift=-.5ex]current bounding box.center)}]
\draw[gray] (0, 0) -- (1, 0);
\node at (0.5, 0) {\normalsize $\uparrow$};
\end{tikzpicture}}= \ket{\frac{}{}\begin{tikzpicture}[scale = 0.5, baseline = {([yshift=-.5ex]current bounding box.center)}]
\draw[gray] (0, 0) -- (1, 0);
\draw[red, fill = red] (0.5,0) ellipse (0.45 and 0.1);
\end{tikzpicture}}$ 
and $\ket{ \begin{tikzpicture}[scale = 0.5, baseline = {([yshift=-.5ex]current bounding box.center)}]
\draw[gray] (0, 0) -- (1, 0);
\node at (0.5, 0) {\normalsize $\downarrow$};
\end{tikzpicture}} = \ket{\frac{}{} \begin{tikzpicture}[scale = 0.5, baseline = {([yshift=-.5ex]current bounding box.center)}]
\draw[gray] (0, 0) -- (1, 0);
\end{tikzpicture}}$.
If we then choose $R_b$ to be slightly greater than $2a$ as shown above, each kagome vertex can neighbor at most one dimer.

\begin{figure}
	\centering
    
    \includegraphics[width=\linewidth]{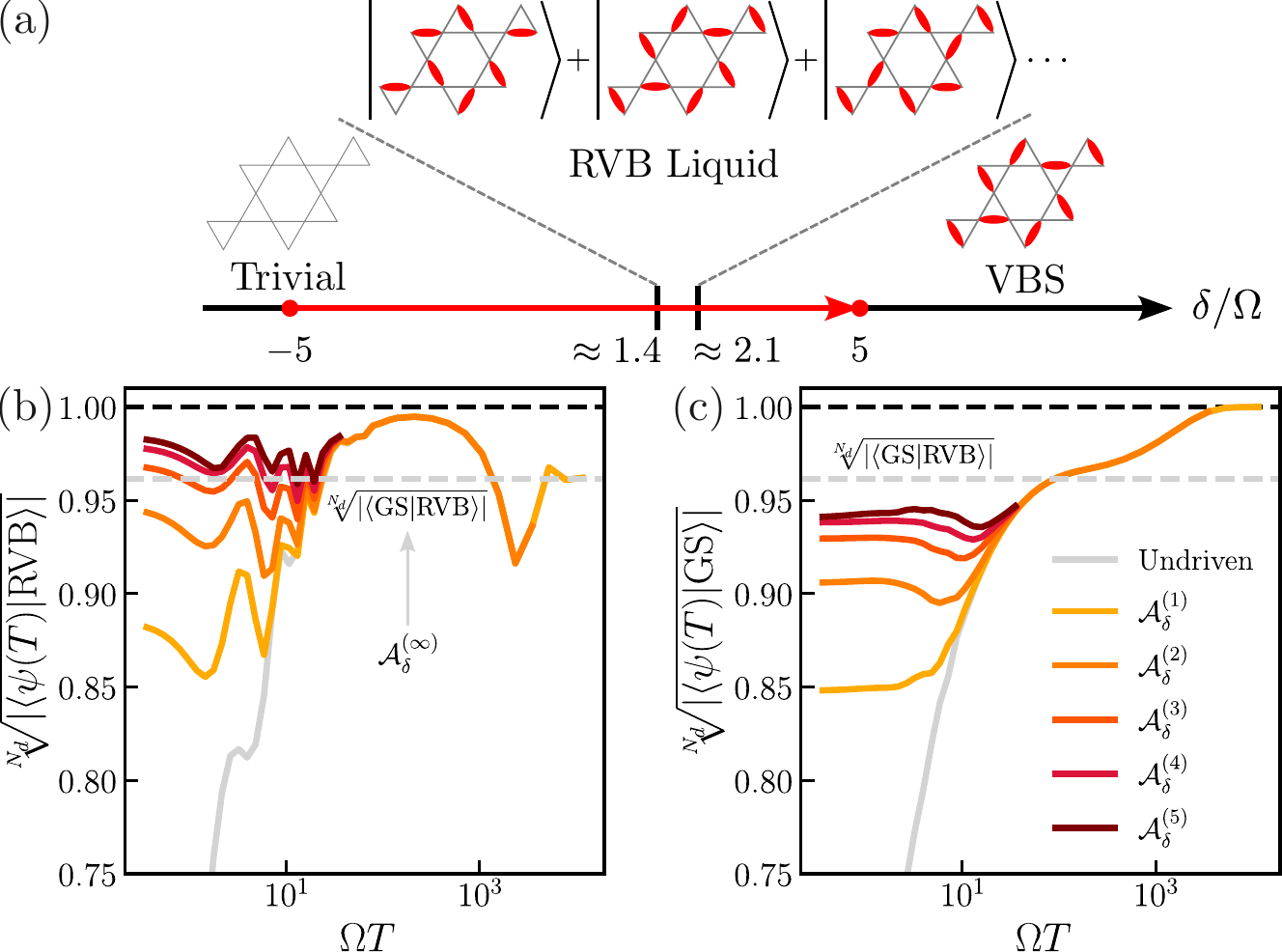}

	\caption{
    \textbf{Shortcuts to Hemidiabaticity in the Rydberg Ruby Lattice.} 
    (a) Starting from the trivial phase of the PXP model of 36 Rydberg atoms on a ruby lattice [Eq.~\eqref{eq:ruby PXP Hamiltonian}], we sweep through the small RVB liquid phase into the VBS phase.
    (b) At hemidiabatic rates ($\Omega T \approx 10^2$), such undriven sweeps (gray) prepare an approximate RVB state, as in previous work \cite{Semeghini_2021,Giudici_2022,Sahay_2023}. 
    As the order of approximate CD driving is increased (color), the state prepared by faster sweeps also approaches the RVB state.
    For experimental implementation, see Fig.~\ref{fig:ruby speedup}.
    (c) Conversely, the VBS ground state is not targeted by approximate AGPs despite exact CD driving (at infinite order) necessarily preparing this state.
    Simulations are performed in the translation and inversion symmetric subspace of dimension $11438 \equiv 2^{N_d}$.
    }
	\label{fig:ruby acd driving}
\end{figure}

Our undriven sweep begins at large negative $\delta$ where the ground state corresponds to an empty state in the trivial or ``Higgs'' phase [Fig.~\ref{fig:ruby acd driving}(a)] \cite{Verresen_2021,Sahay_2023}.
We then sweep to large positive $\delta$ where the ground state has maximized the number of dimers subject to the blockade constraint.
Although there are an exponential number of such dimer coverings of the lattice, only a small subset participates in this valence bond solid (VBS) or ``confined'' phase ground state.
Despite this, 
the dynamically prepared state for experimentally accessible timescales is an approximate resonating valence bond (RVB) state: the equal weight, equal phase superposition of \textit{all} dimer coverings \cite{Semeghini_2021,Giudici_2022,Sahay_2023}.
Indeed, we simulate such an undriven sweep (gray in Fig.~\ref{fig:ruby acd driving})
and find a peak in the RVB overlap at hemidiabatic timescales
even though
the RVB quantum spin liquid phase only occupies a sliver of the phase diagram and is destabilized by tiny perturbations like the van der Waals tails
\cite{Verresen_2021,Semeghini_2021}.

Before elucidating the reason for this phenomenon, let us show that approximate CD driving inherits the same behavior and targets the hemidiabatic regime even beyond first order.
The first order AGP is given by $\mathcal{A}_{\delta}^{(1)}  = -\alpha (\Omega/2) \sum_i P Y_i P$
and is easy to implement experimentally by tuning the phase of the Rabi laser.
At higher orders, it has the form
\begin{align}
    \mathcal{A}_{\delta}^{(\ell)}(\delta) = - \frac{\Omega}{2} \sum_{k=1}^{\ell} \alpha_k(\delta) \underbrace{[H(\delta),\dots [H(\delta)}_{2k-2},  P Y P ]] ,
    \label{eq:ruby agp}
\end{align}
where $PYP = \sum_i P Y_i P$
and the $\alpha_k$ are variationally optimized~\cite{Sels_2017,SM}.
Clearly, all orders beyond $k=1$ are unphysical in the sense that no such terms appear in the native Hamiltonian,
but we will soon show how to realize them by simply modulating the phase of $\Omega$.
The results of driving up to fifth order are shown in Fig.~\ref{fig:ruby acd driving}(b-c).
It is clear that higher-order AGPs continue to target the RVB state rather than the VBS ground state, in line with our qutrit results.

These results can be understood via the existence of a hemidiabatic window in the emergent timescales of the system.
In particular, the two timescales that define the edges of the hemidiabatic regime are understood in terms of two emergent quasiparticles in the RVB phase.
The $e$ excitation corresponds to the removal of a dimer from the RVB state
and therefore has a large energy set by $\delta$, 
whereas the $m$ excitation corresponds to states without an equal superposition of all dimer coverings and therefore has a small energy set by perturbation theory \cite{Sahay_2023,SM}.
This separation of energy scales implies that $e$'s respond quickly to the sweep while $m$'s remain frozen.

Given this understanding, a hemidiabatic sweep from the trivial phase (where $e$'s have proliferated) to the VBS phase (where $m$'s have proliferated) 
allows the $e$'s to equilibrate out while the $m$'s have no time to nucleate.
Like the qutrit, the final state can be approximated by the projection $\mathcal{P}_G$ of the initial state into the dimer-covering subspace---spanned by states with no $e$'s---such that $\tilde{\mathcal{P}}_G \ket{\psi(0)} \approx \ket{\mathrm{RVB}}$ \cite{SM}.
The presence of the phase transition means that this construction cannot remove all $e$'s from the initial state---instead, a dilute density of $e$'s is left over \cite{Zurek_2005,Polkovnikov_2005,Dziarmaga_2005}, leaving finite-size regions of RVB order coined quantum spin lakes \cite{Sahay_2023}.

At low orders, the approximate AGP inherits this behavior and only removes $e$'s without nucleating $m$'s.
At infinite order, we will recover the exact AGP and target the ground state [Fig.~\ref{fig:ruby acd driving}(b)]; however, 
this behavior clearly does not set in even at fifth order. 
These results suggest that the approximate AGP (1) only acts on $e$'s at low orders, driving them out of the state, and (2) eventually acts on both excitations at high enough order, removing $e$'s and nucleating $m$'s.
We will solidify this intuition with the Landau-Ginzburg model.

\begin{figure}[t!]
	\centering

    \includegraphics[width=\linewidth]{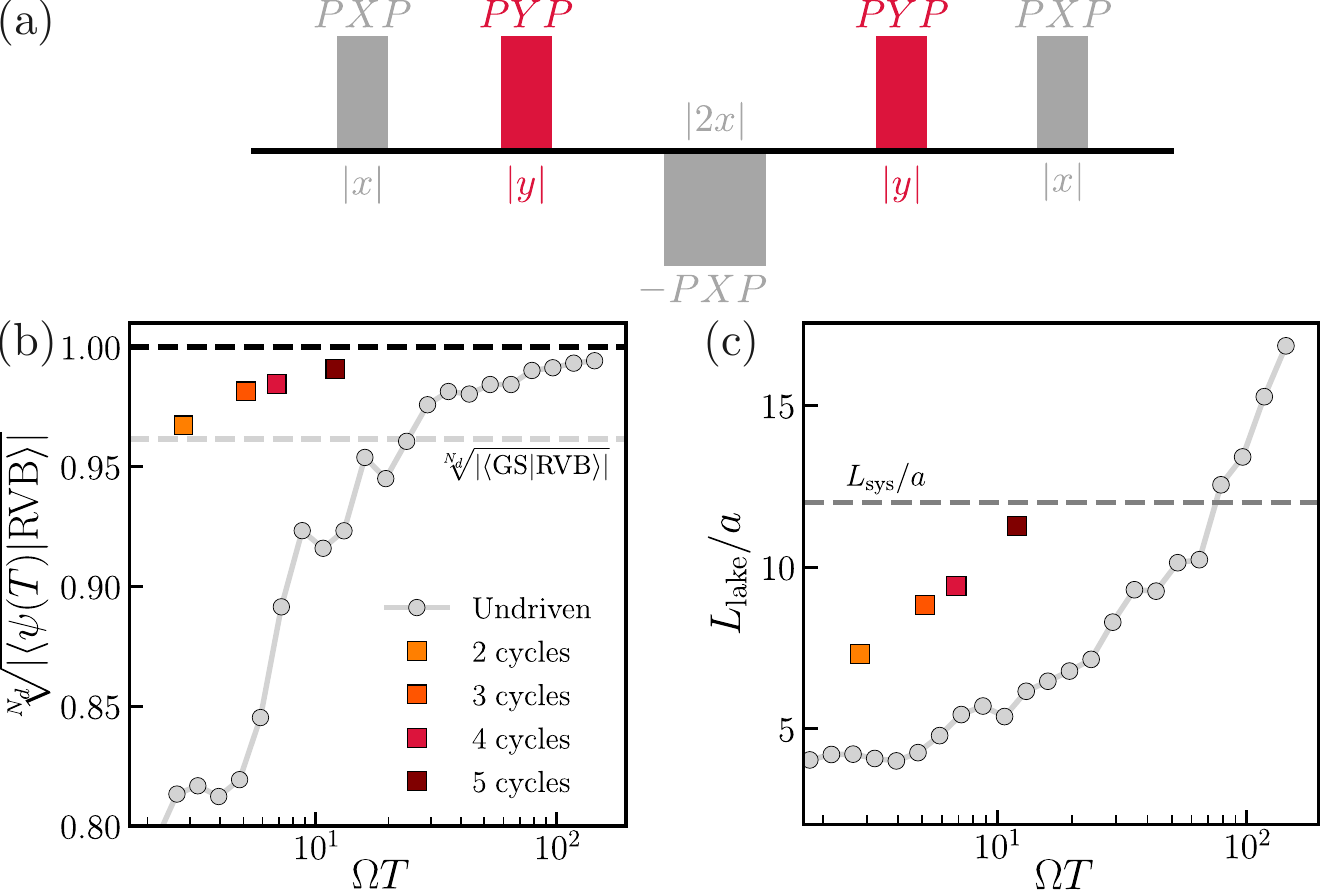}

	\caption{\textbf{Pulse Sequences in the Rydberg Ruby Lattice.} 
    (a) By dressing $PYP$ evolution with $PXP$ pulses, we effectively evolve under terms present in the approximate AGP [Eq.~\eqref{eq:ruby pulse eff H}].
    These ``cycles'' composed of 5 pulses can be concatenated to construct sequences that efficiently prepare an approximate RVB state. 
    (b) We optimize $x,y$ in each cycle to maximize the final dimer density and plot the overlap density with the RVB state as a function of total preparation time.
    Comparing the overlap densities of the states prepared by the pulse sequences with those of the undriven sweep, we find the pulse sequences generate approximate RVB states 
    nearly an order of magnitude faster.
    (c) By estimating the smallest lightcone needed for a circuit to prepare these states (see End Matter), we calculate a ``lake size'' $L_{\mathrm{lake}}$ to quantify the scale over which the prepared states display RVB order and compare it to the system size $L_{\rm sys}$.
    }
	\label{fig:ruby speedup}
\end{figure}

\textbf{Experimental Protocol.}
Although these higher-order approximate AGPs are difficult to implement in experiment, we now consider
a minimal driving sequence that realizes 
such terms
and corresponds to pulsing the Rabi laser at zero detuning for various amounts of time 
[Fig.~\ref{fig:ruby speedup}(a)]:
\begin{align}
    U_{c} = e^{-i x PXP} e^{-i y PYP} e^{2 i x PXP} e^{-i y PYP} e^{-i x PXP} ,
    \label{eq:ruby cycle unitary}
\end{align}
where $U_c$ is repeated for $N_c$ cycles. 
By dressing the first order AGP ($PYP$) with conjugation under $PXP$ unitaries, each cycle implements evolution under the following effective Hamiltonian:
\begin{align}
    2 y \sum_{k=1}^{\infty} \frac{(-ix)^{2k-2}}{(2k-2)!} \underbrace{[PXP,\dots [PXP}_{2k-2},PYP ]] + \mathcal{O}(y^2) .
    \label{eq:ruby pulse eff H}
\end{align}
This operator contains the same terms as the approximate AGP in Eq.~\eqref{eq:ruby agp} when $\delta = 0$.
Even with this restriction, 
this minimal CD-inspired pulse sequence can quickly prepare quantum lakes\footnote{
Optimizing and driving with an AGP ansatz using just the terms in Eq.~\eqref{eq:ruby pulse eff H} yields results very similar to those in Fig.~\ref{fig:ruby acd driving} \cite{SM}.}.

In particular, the system is initialized in the ground state for $\delta = -5\Omega$.
Then, for a given $N_c$, all $2N_c$ values of $x$ and $y$ 
are optimized such that the density of dimers in the final state 
is maximized (similar to \cite{duncan2025counterdiabaticinfluencedfloquetengineeringstatepreparation}), heuristically targeting $e$'s in the initial state.
Crucially, Fig.~\ref{fig:ruby speedup}(b) shows that the resulting state has a high overlap with the RVB state.
Remarkably, these fidelities can be achieved nearly an order of magnitude more quickly than in the undriven case!
Such a speedup relative to the coherence time opens the door to performing experiments that go beyond the preparation of the spin lakes state.
Moreover, Fig.~\ref{fig:ruby speedup}(c) shows that the 
lake size $L_{\rm lake}$ 
quickly approaches the system size $L_{\rm sys}$ as we increase $N_c$.
While 
$L_{\rm lake}$ can be heuristically interpreted as the average distance between $e$'s in the final state, the precise definition 
can be found in the End Matter.

One might worry that the performance of our optimized protocols will suffer when applied to larger experimental systems with perturbations such as van der Waals tails. 
Indeed, CD driving generally trades long, robust sweeps for short protocols that are sensitive to perturbations \cite{SM,Gjonbalaj_2022}.
In the SM \cite{SM}, we provide preliminary evidence that our optimized protocols still perform well as the system size increases; furthermore, one can use on-device optimization to maximize performance even with such perturbations, as recently shown in Ref.~\cite{geim2026engineeringquantumcriticalitydynamics}.

\begin{figure}[t!]
	\centering

    \includegraphics[width=\linewidth]{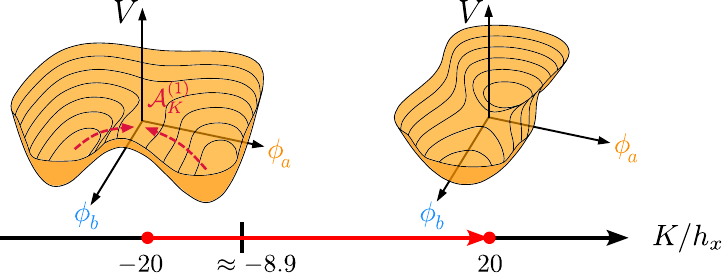}

	\caption{\textbf{Hemidiabaticity in a Landau-Ginzburg Model.} 
    We show the potential [Eq.~\eqref{eq:lg potential}] of a two-component Landau-Ginzburg model that provides a semi-classical picture of hemidiabaticity.
    We sweep $K$ from a regime in the model where $\langle\phi_a\rangle \neq 0$ to one where $\langle\phi_b\rangle \neq 0$.
    $\mathcal{A}_K^{(1)}$ (dashed arrows) acts as a state-dependent force that sends $\phi_a \to 0$ without affecting $\phi_b$.
    We show numerical evidence for this model's hemidiabatic regime in the End Matter.
    }
	\label{fig:full LG diagram}
\end{figure}

\textbf{Generalizations and Semi-Classical Picture.} 
We now turn to a
Landau-Ginzburg effective field theory 
that both distills many-body hemidiabaticity 
to its essential ingredients and provides semi-classical intuition for the driving's effectiveness.
This field theory's Hamiltonian density is $\mathcal{H} =  \frac{1}{2} \left( \Delta_a \Pi_a^2 + \Delta_b \Pi_b^2 + f_a (\nabla \phi_a)^2 + f_b (\nabla \phi_b)^2 \right)  + V$,
where $\phi_a$ ($\phi_b$) is a scalar field describing boson $a$ ($b$).
$\Pi_i$ is the canonical field momentum,
and
$\Delta_i$ ($f_i$) quantifies the dynamical timescale (spatial coupling) of each field.
The potential $V$ (Fig.~\ref{fig:full LG diagram}) is 
\begin{align}
    V = \frac{1}{2} K \phi_a^2 + \frac{\lambda_a}{4!} \phi_a^4 + \frac{1}{2} (h_x \phi_a^2 - h_z) \phi_b^2 + \frac{\lambda_b}{4!} \phi_b^4 ,
    \label{eq:lg potential}
\end{align}
where we have used the same parameter labels as the qutrit for conceptual clarity.
Although the qutrit's 
timescales 
follow from $K,h_x,$ and $h_z$, we are able to tune the $\Delta_i$ values independently in this more general model and choose $\Delta_a \gg \Delta_b$ such that $a$'s are fast and $b$'s are slow.
This effective field theory is independent of any specific microscopic origin
and provides an alternative, intrinsically many-body perspective on hemidiabaticity.

The model 
has two dominant equilibrium phases: when $K\ll0$,
the $a$'s form a Bose-Einstein condensate ($\langle \phi_a \rangle \neq 0$), analogous to how the $e$'s proliferate in the trivial phase of the Rydberg model; when $K\gg0$ ,
the $b$'s form a condensate ($\langle \phi_b \rangle \neq 0$),
analogous to how the $m$'s proliferate in the VBS phase. 
We show in the End Matter that, as expected, choosing the model parameters to separate the bosons' 
timescales causes a hemidiabatic window to emerge in the sweep dynamics where \textit{neither} boson has condensed.
Furthermore, the first order AGP, given by $\mathcal{A}^{(1)}_K \propto - \int d^2 x \phi_a \Pi_a$, 
extends this window
by removing $a$'s without nucleating $b$'s.
However, in this field theory, the AGP takes on an intriguing semi-classical interpretation as a \textit{state-dependent force}.
In particular, if $\phi_a > 0$, $\Pi_a$ translates the field value $\phi_a \to 0$.
Similarly, if $\phi_a < 0$, $\Pi_a$ translates in the opposite direction and $\phi_a \to 0$ still holds. 
The AGP is therefore able to uncondense the $\phi_a$ field without touching $\phi_b$ (Fig.~\ref{fig:full LG diagram}), a behavior that we argue extends to higher orders in the End Matter.

\textbf{Acknowledgements.} We thank 
M. Bukov, P. Dolgirev, R. Fan, A. Gu, N. U. Köylüoğlu, F. Machado, N. Maskara, S. Morawetz, S. Ostermann, A. Polkovnikov, M. Serbyn, M. Szurek, and R. Verresen for fruitful discussions.
We especially thank F. Machado, A. Polkovnikov, and R. Verresen for a careful reading of this manuscript.
We thank Harvard University's FAS Research Computing for numerical resources. 
N.O.G. acknowledges support from the Generation Q G2 Fellowship.
R.S. acknowledges support from the U.S. Department of Energy, Office of Science, Office of Advanced Scientific Computing Research, Department of Energy Computational Science Graduate Fellowship under Award Number DESC0022158.
S.F.Y. acknowledges NSF via PHY-2207972, the CUA PFC PHY-2317134 and the HDR Q-IDEAS grant OAC-2118310.

\textcopyright~2026 American Physical Society. 
This is the accepted manuscript of the following article:
N.~O. Gjonbalaj, R.~Sahay, and S.~F. Yelin,
“Shortcuts to Analog Preparation of Nonequilibrium Quantum Lakes”,
Phys. Rev. Lett. 
\textbf{137}, 
090802
(2026).
The final published version is available from \href{https://doi.org/10.1103/7bc2-6cgy}{https://doi.org/10.1103/7bc2-6cgy}. 
This manuscript version is posted with permission for non-commercial scholarly use.


%

\clearpage
\newpage
\section*{End matter}

\subsection*{Defining the lake size}
While the overlap density confirms that we approximately prepare the RVB state in the Rydberg system, a more physical question to ask is over what length scale the state is indistinguishable from this target.
In other words, how large is the quantum spin lake?
As an estimate, we use the lightcone size of the minimum depth circuit needed to prepare the lakes state.
In particular, we compute an estimate of the minimal circuit depth $D_{\mathrm{min}}$ required to prepare a state with the final density of $e$ and $m$ excitations after the pulse sequence \cite{tikku2022circuitdepthversusenergy}.
The heuristic definitions of each excitation from the text can be made exact using the loop operators from \cite{Verresen_2021,Semeghini_2021,Sahay_2023}.
More specifically, the dimer covering constraint, referred to as the Gauss law, is defined using the operator
\begin{equation}\label{eq-RydbergGauss}
G_v = \begin{tikzpicture}[scale = 1.2, baseline={([yshift=-.5ex]current bounding box.center)}]
\draw[gray] (0,0) -- ({-0.5 * 1/2}, {0.5 * 0.866025404}) -- ({0.5 * 1/2}, {0.5 * 0.866025404}) -- cycle;
\draw[gray] (0,0) -- ({-0.5 * 1/2}, {-0.5 * 0.866025404}) -- ({0.5 * 1/2}, {-0.5 * 0.866025404}) -- cycle;
\draw[orange(ryb), dashed, line width = 0.4 mm] (0,0) circle (7 pt);
\end{tikzpicture}  \quad , \quad
\begin{tikzpicture}[scale = 1.2, baseline={([yshift=-.5ex]current bounding box.center)}]
\draw[gray] (0,0) -- ({-1.5*(0.5) * 1/2}, {-(1.5*0.5) * 0.866025404}) -- ({(1.5*0.5) * 1/2}, {-(1.5*0.5) * 0.866025404}) -- cycle;
\draw[orange(ryb), dashed, line width = 0.4 mm] (-1.5*0.3, 1.5*-0.25) -- (1.5*0.3, 1.5*-0.25);
\end{tikzpicture}\ \ =\ \  
\begin{tikzpicture}[scale = 1.2, baseline={([yshift=-.5ex]current bounding box.center)}]
\draw[gray] (0,0) -- ({-1.5*(0.5) * 1/2}, {-(1.5*0.5) * 0.866025404}) -- ({(1.5*0.5) * 1/2}, {-(1.5*0.5) * 0.866025404}) -- cycle;
\node at ({-1.5*(0.5) * 1/4}, {-(1.5*0.5) * 0.866025404/2}) {\normalsize $Z$};
\node at ({-1.5*(0.5) * 1/4 + 0.75/2}, {-(1.5*0.5) * 0.866025404/2}) {\normalsize $Z$};
\end{tikzpicture} .
\end{equation}
When $G_v = -1$, the vertex has an odd number of neighboring atoms in the Rydberg state. In the PXP approximation, this number must be 1, implying that the dimer covering constraint is at least satisfied locally. When $G_v = -1$ for all vertices, the state is fully packed with dimers. On the other hand, $G_v = 1$ means the vertex has an even number of neighboring Rydberg states, and in the blockaded subspace this even number must be 0. The dimer covering constraint is violated locally and indicates the presence of an $e$ excitation.
This definition allows us to explicitly write the projector $\mathcal{P}_G = \prod_v (1 - G_v)/2$ and confirm that the projection formula, which claims that the RVB state is given by projecting $e$ excitations out of the initial ground state, holds to an error $1 - |\langle \mathrm{RVB} | \tilde{\mathcal{P}}_G | \psi(0) \rangle | \approx 5.3 \times 10^{-12}$ in our system.

Likewise, the $m$ excitation is defined locally using the operator
\begin{equation} \label{eq-RydbergWilson}
W_{p} = \begin{tikzpicture}[scale = 1, baseline={([yshift=-.5ex]current bounding box.center)}]
\draw[gray] (0,0) -- ({-0.5 * 1/2}, {0.5 * 0.866025404}) -- ({0}, {0.866025404}) -- ({0.5}, {0.866025404}) -- ({ 0.5 +0.5 * 1/2}, {0.5 * 0.866025404}) -- (0.5, 0) -- cycle;
\draw[dodgerblue,decorate, decoration={snake, segment length=1.5mm}, line width = .35mm] (0,0) -- ({-0.5 * 1/2}, {0.5 * 0.866025404});
\draw[dodgerblue,decorate, decoration={snake, segment length=1.5mm}, line width = .35mm] ({-0.5 * 1/2}, {0.5 * 0.866025404}) -- ({0}, {0.866025404});
\draw[dodgerblue,decorate, decoration={snake, segment length=1.5mm}, line width = .35mm] ({0}, {0.866025404}) --  ({0.5}, {0.866025404});
\draw[dodgerblue,decorate, decoration={snake, segment length=1.5mm}, line width = .35mm] ({0.5}, {0.866025404}) -- ({ 0.5 +0.5 * 1/2}, {0.5 * 0.866025404});
\draw[dodgerblue,decorate, decoration={snake, segment length=1.5mm}, line width = .35mm] ({ 0.5 +0.5 * 1/2}, {0.5 * 0.866025404}) -- (0.5, 0);
\draw[dodgerblue,decorate, decoration={snake, segment length=1.5mm}, line width = .35mm] (0.5, 0) -- (0,0);
\end{tikzpicture}
\quad ,\quad
\begin{tikzpicture}[scale = 1.2, baseline={([yshift=-.5ex]current bounding box.center)}]
\draw[gray] (0,0) -- ({-1.5*(0.5) * 1/2}, {-(1.5*0.5) * 0.866025404}) -- ({(1.5*0.5) * 1/2}, {-(1.5*0.5) * 0.866025404}) -- cycle;
\draw[dodgerblue,decorate, decoration={snake, segment length=1.5mm}, line width = .35mm] ({-1.5*(0.5) * 1/2}, {-(1.5*0.5) * 0.866025404}) -- ({(1.5*0.5) * 1/2}, {-(1.5*0.5) * 0.866025404});
\end{tikzpicture}  = \begin{cases}
\begin{tikzpicture}[scale = 1.2, baseline={([yshift=-.5ex]current bounding box.center)}]
\draw[gray] (0,0) -- ({-.7*(0.5) * 1/2}, {-(.7*0.5) * 0.866025404}) -- ({(.7*0.5) * 1/2}, {-(.7*0.5) * 0.866025404}) -- cycle;
\draw[red, fill = red, rotate around={60:({-.35*(0.5) * 1/2}, {-(.35*0.5) * 0.866025404})}] ({-.35*(0.5) * 1/2}, {-(.35*0.5) * 0.866025404}) ellipse (0.175 and 0.03);
\end{tikzpicture} \leftrightarrow
\begin{tikzpicture}[scale = 1.2, baseline={([yshift=-.5ex]current bounding box.center)}]
\draw[gray] (0,0) -- ({-.7*(0.5) * 1/2}, {-(.7*0.5) * 0.866025404}) -- ({(.7*0.5) * 1/2}, {-(.7*0.5) * 0.866025404}) -- cycle;
\draw[red, fill = red, rotate around={-60:({(.35*0.5) * 1/2}, {-(.35*0.5) * 0.866025404})}]  ({(.35*0.5) * 1/2}, {-(.35*0.5) * 0.866025404})  ellipse (0.175 and 0.03);
\end{tikzpicture}
\\
\begin{tikzpicture}[scale = 1.2, baseline={([yshift=-.5ex]current bounding box.center)}]
\draw[gray] (0,0) -- ({-.7*(0.5) * 1/2}, {-(.7*0.5) * 0.866025404}) -- ({(.7*0.5) * 1/2}, {-(.7*0.5) * 0.866025404}) -- cycle;
\end{tikzpicture}
\leftrightarrow
\begin{tikzpicture}[scale = 1.2, baseline={([yshift=-.5ex]current bounding box.center)}]
\draw[gray] (0,0) -- ({-.7*(0.5) * 1/2}, {-(.7*0.5) * 0.866025404}) -- ({(.7*0.5) * 1/2}, {-(.7*0.5) * 0.866025404}) -- cycle;
\draw[red, fill = red] ({0}, {-(.7*0.5) * 0.866025404}) ellipse (0.175 and 0.03); 
\end{tikzpicture} \: .
\end{cases}
\end{equation}
This operator flips spins on kagome triangles that participate in the plaquette in such a way that dimer coverings are mapped to each other. Measuring $W_p = 1$ therefore implies that all such dimer coverings have equal weight and phase, and locally the superposition has the correct structure for the RVB state. If $W_p = -1$, however, some phases are $-1$ and the superposition does not have the correct local structure. In this case, an $m$ excitation lives on the plaquette.

These operators allow us to define the following local stabilizer Hamiltonian:
\begin{align}
    H_{\mathrm{stab}} = \sum_v \frac{1 + G_v}{2} + \sum_p \frac{1 - W_p}{2} .
\end{align}
Unlike the Rydberg Hamiltonian in the text, the RVB state is one of the zero-energy topological ground states of this Hamiltonian, with the ground state degeneracy depending on the topology of the lattice.
A finite energy density therefore quantifies the remaining density of excitations above the RVB state.

Given a state with energy density $\epsilon \equiv \langle H_{\rm stab} \rangle / N_q$, where $N_q$ is the number of qubits, the bound from Ref.~\cite{tikku2022circuitdepthversusenergy} states that the depth $D$ of any circuit used to prepare the state must scale as $D = \Omega(\min(1/\epsilon^{\frac{1-\alpha}{2}},\sqrt{N_q}))$ for any $\alpha > 0$. 
Although Ref.~\cite{tikku2022circuitdepthversusenergy} derives this result in the toric code \cite{Kitaev_2003}, the bound also applies to our Rydberg stabilizer Hamiltonian.
Calculating $D_{\mathrm{min}} = 1/\sqrt{\epsilon}$ therefore provides a lower estimate of the depth needed to prepare the states at the end of each pulse sequence from the main text.
To convert this circuit depth into a lake size, we need to quantify the length scale over which a ``gate'' would act in such a circuit. 
We use the interaction range $R_b = 2a$ set by the blockade as an estimate of this scale, giving a final value of $L_{\mathrm{lake}} = R_b D_{\mathrm{min}} = 2a/\sqrt{\epsilon}$ for the lake size.

In the text, we note that quantum spin lakes can intuitively be thought of as the finite-size regions between the $e$ excitations left over after a sweep or driving protocol.
To make contact with this interpretation, let us first note that every unit cell of the kagome lattice hosts 1 plaquette and 3 vertices. 
As such, the $W_p$ contribution to $H_{\mathrm{stab}}$ is suppressed relative to that of $G_v$. 
If we therefore approximate $W_p$ as 1 and neglect this term from the energy density, the lake size definition reduces to $L_{\mathrm{lake}} \to \sqrt{2}(2a/\sqrt{(1 + \langle G_v \rangle)/2})$ by translational invariance.
The expression in parentheses is exactly the average distance between $e$ excitations in the final state.
In particular, the density of $e$'s is given by $n_e = (1 + \langle G_v \rangle)/2$ excitations per vertex. 
Using the distance $2a$ between adjacent vertices, this average distance is indeed given by $L_e \equiv 2a/\sqrt{n_e}$.
In the empty product state $n_i = 0$, this returns $L_e = 2a$, so no lake includes more than a single vertex and the state exhibits no spin liquid correlations. 
However, for the true RVB state, $L_e \to \infty$, indicating that no $e$ excitation will be present even in an infinite system. 
When $L_e > L_{\mathrm{sys}}$, there is, on average, less than one $e$ excitation present in the system.

\subsection*{Landau-Ginzburg dynamics}

\begin{figure}[t!]
	\centering

    \includegraphics[width=0.65\linewidth]{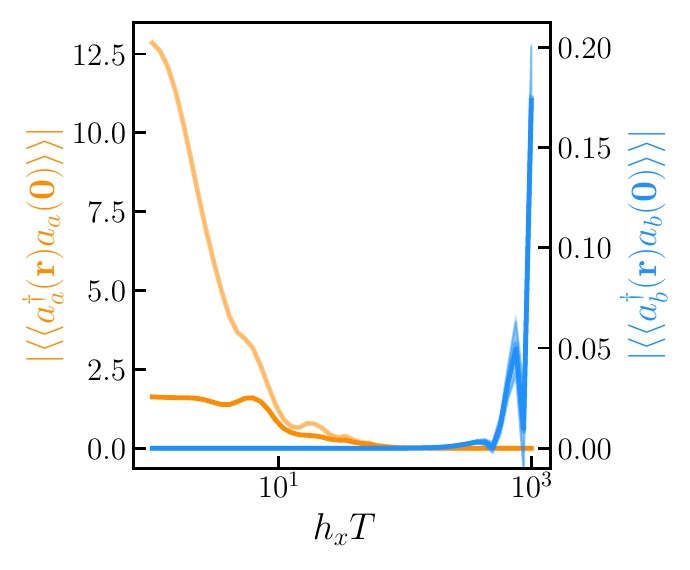}

	\caption{\textbf{Approximate CD Driving in a Landau-Ginzburg Model.} 
    We plot the long-range two-point functions [Eq.~\eqref{eq:lg two point function} 
    with $\mathbf{r}$ set to $(5,0)$]
    ] for $a$ (orange) and $b$ (blue) bosons at the end of an undriven sweep (light color) 
    on a $10 \times 10$ periodic lattice
    to quantify how much each boson has condensed.
    A clear hemidiabatic window appears ($h_x T \approx 10^2$) where neither boson has condensed, and since $\mathcal{A}_K^{(1)}$ only acts on $\phi_a$, approximate CD driving extends this window to faster sweeps (dark color).
    The ribbons indicate the standard error of the mean for each curve.
    }
	\label{fig:full LG numerics}
\end{figure}

We now confirm that a hemidiabatic window exists in the sweep dynamics of the Landau-Ginzburg model from the main text.
To do so, we use the truncated Wigner approximation \cite{Blakie_2008,Polkovnikov_2010,WOOTTERS19871,Fano_1957} to simulate a sweep from $K = -20 h_x$ to $K = 20 h_x$ on a two-dimensional 
$10 \times 10$ periodic lattice of oscillators
and then subsequently examine the condensation of $\phi_{a,b}$ in the final state.
We probe this condensation via the long-range component of the  two-point function: 
\begin{align}
    \lim_{\mathbf{r}\to\infty} \langle a^{\dagger}_{a, b}(\mathbf{r}) a_{a, b}(\mathbf{0}) \rangle ,
    \label{eq:lg two point function}
\end{align}
which probes the magnitude of $\langle a^{\dagger}_{a, b}(\mathbf{0}) \rangle$~\cite{altland2010condensed, sahay2024superconductivity,SM}.
To probe long-range correlations without wrapping around the torus, we choose $\mathbf{r} = (5,0)$.
We define $a_a \propto \phi_a + i \Pi_a/\sqrt{K(T)/\Delta_a}$ ($a_b \propto \phi_b + i \Pi_b/\sqrt{h_z/\Delta_b}$) as the  annihilation operator of $\phi_a$ ($\phi_b$) on a single site with the potential
\begin{align}
    V_{\rm site} = \frac{1}{2} K(T) \phi_a^2 + \frac{1}{2} h_z \phi_b^2 .
\end{align}
This choice of potential corresponds to a phase where neither field is condensed in the ground state, analogous to the spin liquid phase in the Rydberg setting.
As such, $a$ and $b$ bosons are well defined as excitations above this equilibrium and furthermore quantify how far a prepared state is from the uncondensed target state.

The results of this simulation are pictured in Fig.~\ref{fig:full LG numerics}.
By choosing $\Delta_a \gg \Delta_b$, the undriven dynamics (shown in light color) have a clear hemidiabatic window near $h_xT = 10^2$ where neither boson is present, a state inaccessible in equilibrium. 
To gain a more quantitative understanding of the separation of timescales in this model, consider expanding the Hamiltonian density around the mean field solution for large negative $K$:
\begin{align}
    \mathcal{H}_{\rm mf} (K \ll 0) \to 
    \frac{1}{2}  \Big( &\Delta_a \Pi_a^2 + \Delta_b \Pi_b^2 \nonumber \\
    - & 2 K \delta \phi_a^2 - \left(\frac{6 h_x K}{\lambda_a} + h_z \right) \delta \phi_b^2 \Big)
\end{align}
where we have dropped additive constants and $\delta \phi_i = \phi_i - \langle \phi_i \rangle_{\rm mf}$ is the deviation of $\phi_i$ from its mean field solution.
By treating each boson's contribution as a single quantum harmonic oscillator, it is clear that the oscillation frequency is given by $\sqrt{-2 \Delta_a K}$ for the $a$ bosons and $\sqrt{-\Delta_b(6 h_x K/\lambda_a + h_z)}$ for the $b$ bosons.
Thus, as $K \to - \infty$, the ratio of these frequency scales is determined by $\sqrt{\Delta_a/\Delta_b}$.
This is analogous to the qutrit, where both excited states have similar gaps above the ground state for large negative $K$ (see Fig.~\ref{fig:qutrit}(a)).

Similarly, when $K$ is large and positive, the mean field expansion of the Hamiltonian density is
\begin{align}
    \mathcal{H}_{\rm mf} (K \gg 0) \to 
    \frac{1}{2}  \Big( &\Delta_a \Pi_a^2 + \Delta_b \Pi_b^2 \nonumber \\
    + & \left( K + \frac{6 h_x h_z}{\lambda_b} \right) \delta \phi_a^2 + 2 h_z \delta \phi_b^2 \Big)
\end{align}
and the oscillation frequency becomes 
$\sqrt{\Delta_a (K + 6 h_x h_z/\lambda_b)}$ ($\sqrt{2 \Delta_b h_z}$)
for the $a$ ($b$) bosons.
At a large positive value of $K$, the ratio of these scales is proportional to $\sqrt{\Delta_a K / \Delta_b h_z}$.
Again, this agrees with the qutrit, where the ratio of the two excited state gaps grows as $K > 0$ increases.
It is therefore clear that choosing $\Delta_a \gg \Delta_b$ ensures that $a$ bosons will have a much faster inherent timescale than $b$ bosons on both sides of the transition.
Although one can choose $\Delta_i(K)$ to have the functional dependence to match the gaps seen in Fig.~\ref{fig:qutrit}, we instead choose constant $\Delta_i$ to simplify the sweep and driving.

As explained in the text, approximate CD driving still targets the hemidiabatic regime in this model. 
At leading order, the AGP takes the form $\mathcal{A}^{(1)}_K = \Delta_a \alpha \int d^2 x  \phi_a \Pi_a$, where $\alpha$ is calculated numerically \cite{SM}. 
The driven dynamics are shown in Fig.~\ref{fig:full LG numerics} in dark color. 
The driving clearly only affects the $a$ bosons---reducing the density of $a$'s without affecting the $b$'s---and widening the hemidiabatic window.
As we include higher order contributions to the AGP, we will eventually drive $b$ dynamics and approach the true ground state rather than the lakes state. 
Indeed, at second order, the AGP already includes terms like $\Delta_b h_x \phi_a^2 \phi_b \Pi_b$. 
However, such terms are limited by the rate ($\Delta_b$) and energy scale ($h_z$) of $b$ nucleation relative to other terms at the same order targeting the $a$ sector.
One might worry that optimizing the driving could yield a very large $\alpha$ at this order (e.g. scaling as $1/\Delta_b^2$) and invalidate the argument.
However, the presence of other terms at the same order in the AGP which do not contain $\Delta_b,h_z$ (such as terms targeting the field $\phi_a$) eliminates this loophole. $\alpha$ must be optimized for all terms at the same order simultaneously; as such, the $b$ terms get drowned out before $\alpha$ is even chosen.
Thus, the AGP remains hemidiabatic even after optimizing the $\alpha$ coefficients.
We argue that these ingredients---the separation of scales between the $a$ bosons ($K,\Delta_a$) and $b$ bosons ($h_z,\Delta_b$)---are sufficient for a system to host a hemidiabatic regime.

\clearpage
\newpage
\onecolumngrid
\appendix

\begin{center}
{\large \textbf{Supplemental Materials}}    
\end{center}

\section{Supplemental Materials A: Interpretation and Review}

\subsection{Advantages of hemidiabatic sweeps}

Although we have argued that hemidiabatic sweeps allow for the creation of quantum lakes when the desired order does not appear in the ground state, one might ask what advantage this method offers when the order \textit{does} exist in the ground state. Indeed, approximate CD driving can be used to accelerate such sweeps and avoid the issue of slow defects (e.g. $m$ excitations in the Rydberg case or $b$ bosons in the Landau-Ginzburg case) altogether.

In the scenarios we have outlined, where the gap remains small for a portion of the parameter sweep, such an ordered phase will typically only occupy a tiny sliver of the phase diagram. This occurs because the phase is only protected by a small gap, so small perturbations can drive the system into a nearby disordered phase. As such, the sweep must be fine-tuned to land exactly within this tiny sliver. The hemidiabatic construction, in contrast, opens up a large portion of the phase diagram for state preparation. Disordered phases where the slow defects have proliferated are still able to prepare the ordered phase, removing much of the fine-tuning necessary in the adiabatic case.

In addition, the small gap in the ordered phase generically causes the correlation length $\xi$ to be large (see e.g. Ref.~\cite{Verresen_2021} for the correlation length in the PXP case). Thus, even if one successfully sweeps adiabatically into the ordered phase, the resulting order is difficult to probe locally; one must coarse-grain the system to verify the desired order has been reached \cite{Cong_2024}. In contrast, hemidiabatic sweeps allow for the creation of lakes which are large compared to the correlation length and can approach the system size: $\xi \ll L_{\mathrm{lake}} \sim L_{\mathrm{sys}}$. With this hierarchy, large-scale order is more easily detectable than the adiabatic case, where scales like $\xi \lesssim L_{\mathrm{sys}}$ can easily shroud the order.

\subsection{Approximate counterdiabatic driving}
\label{sec:acd driving}

We now review counterdiabatic 
driving and its variational approximation \cite{Demirplak_2003,Demirplak_2005,Berry_2009,Kolodrubetz_2017,Sels_2017}. Consider a Hamiltonian $H(K(t))$ as a function of $K$, a parameter which we can sweep.
We can understand how diabatic transitions occur during this sweep by going to the moving frame where $H$ is instantaneously diagonalized. This transformation is implemented by a unitary $U(K)$ which explicitly depends only on the parameter $K$. In this new frame, the effective Hamiltonian acquires a new term similar to e.g. the Coriolis and centrifugal forces:
\begin{align}
    \tilde{H}^{\mathrm{eff}} = \tilde{H} - \dot{K} \tilde{\mathcal{A}}_K ,
\end{align}
where the tilde indicates the moving frame, e.g. $\tilde{H} = U^{\dagger} H U = \sum_n E_n \ket{n(K)} \bra{n(K)}$. The object $\tilde{\mathcal{A}}_K$ is known as the ``adiabatic gauge potential'' or AGP 
and is the infinitesimal generator of the unitary $U(K)$:
\begin{align}
    \tilde{\mathcal{A}}_K = i U^{\dagger} \partial_{K} U .
\end{align}
Since $\tilde{H}$ is by definition diagonal, the source of diabatic transitions is the AGP term, and as expected, the magnitude of this term grows as the speed of the sweep grows. However, the moving frame Hamiltonian clearly motivates a method for exactly following any eigenstate at any sweep rate. Consider evolving under the following Hamiltonian in the lab frame:
\begin{align}
    H_{\mathrm{CD}} = H + \dot{K} \mathcal{A}_K ,
\end{align}
where $\mathcal{A}_K = U \tilde{\mathcal{A}}_K U^{\dagger}$ is the AGP in the lab frame. When this counterdiabatic Hamiltonian is expressed in the moving frame, the AGP terms cancel and the resulting Hamiltonian is completely diagonal. Therefore, the $\dot{K} \mathcal{A}_K$ term applies a driving in the lab frame which exactly cancels off diabatic transitions.

However, the AGP is only well defined when all eigenstates change continuously as we vary $K$. In systems which are integrable or have well-separated levels, this condition is met and CD driving can be implemented relatively easily. However, chaotic systems and gapless points will have AGPs which blow up, reflecting the fact that eigenstates are extremely sensitive to perturbations. This effect can be seen more clearly if we write the off-diagonal matrix elements of the AGP:
\begin{align}
    \bra{m} \mathcal{A}_K \ket{n} = -i \frac{\bra{m} \partial_K H \ket{n}}{E_m - E_n} ,
    \label{eq:agp matrix elements}
\end{align}
an expression familiar from perturbation theory. When a system's gap closes, the denominator vanishes and causes the entire expression to blow up. One can show that a similar divergence occurs for chaotic systems \cite{Kolodrubetz_2017,Pandey_2020}.
Moreover, the exact AGP will not only have diverging magnitude but also highly nonlocal terms present, reflecting how long-range correlations and structure within eigenstates can dramatically change near gap closures.

Motivated by these limitations, the authors of Ref.~\cite{Sels_2017} developed a method for approximating the AGP using a variational procedure. Given some (preferably local) ansatz $\mathcal{A}_K^*$ for the AGP, we consider the operator
\begin{align}
    G_K(\mathcal{A}_K^*) = \partial_K H + i \left[ \mathcal{A}_K^*, H  \right] .
\end{align}
In systems with a finite local Hilbert space dimension, the action for this variational procedure is then calculated as
\begin{align}
    \mathcal{S}(\mathcal{A}_K^*) = \Tr [G_K^2] .
    \label{eq:trace action}
\end{align}
As noted in Ref.~\cite{Sels_2017}, the fact that Pauli matrices are traceless and square to the identity means that such actions can be calculated in spin-1/2 systems analytically. One can then vary the parameters within the AGP ansatz until the action is extremized:
\begin{align}
    \frac{\delta \mathcal{S}}{\delta \mathcal{A}_K^*} = 0 .
\end{align}
Such an approximate AGP represents the best local approximation to the true AGP, and as a result, it can be used in CD driving to eliminate most transitions away from the desired state. This method has been generalized and applied to many models since its inception, including spins, oscillators, their classical counterparts, and fermions \cite{Sels_2017,Claeys_2019,bukov2018reinforcement,hartmann2019rapid,hegade2021shortcuts,hegade2022portfolio,Gjonbalaj_2022,barone2024counterdiabatic}.

Although this method can optimize a given ansatz, it does not suggest how to choose the ansatz itself. To address this, the authors of Ref.~\cite{Claeys_2019} introduced the following form for the AGP:
\begin{align}
    \mathcal{A}_K^{(\ell)} = i \sum_{k = 1}^{\ell} \alpha_k \underbrace{[H,[H,\dots[H}_{2k-1},\partial_K H]]] .
    \label{eq:agp ansatz}
\end{align}
If we send $\ell \to \infty$, the terms will span the entire space of operators present in the exact AGP, and therefore the variational procedure will return the exact form. Assuming $H$ is local, successive terms in the expansion represent more and more nonlocal contributions to the AGP and therefore allow for a controlled treatment of systems where the exact AGP may not exist. Although this ansatz was motivated in various ways in \cite{Claeys_2019}, we are concerned with two in particular. 
First, the off-diagonal elements of this operator are 
\begin{align}
    \bra{m} \mathcal{A}_K^{(\ell)} \ket{n} = i \sum_{k = 1}^{\ell} \alpha_k \omega_{mn}^{2k-1} \bra{m} \partial_K H \ket{n} ,
\end{align}
where $\omega_{mn} = E_m - E_n$. Comparing this with the form in Eq.~\eqref{eq:agp matrix elements}, we see that this ansatz is optimizing $\alpha_k$ to make the best polynomial fit to the function $-1/\omega_{mn}$. Importantly, this fit need only match this function for values of $\omega_{mn}$ which appear in $H$'s spectrum, and as a result different systems will have different optimal parameters $\alpha_k$. 
Crucially, Ref.~\cite{Claeys_2019} notes that all terms in this expansion vanish as $\omega_{mn} \to 0$ and therefore cannot capture the exact divergence there. This ansatz therefore inherently cancels off large transitions more than small ones.
(Although the error of the approximate AGP is also large for very high-energy transitions, these are generally  negligible even without driving. For more detailed analysis, see e.g. \cite{morawetz2025universalcounterdiabaticdriving,finzgar2025counterdiabaticdrivingperformanceguarantees}). As such, it is particularly well-suited for accelerating the preparation of quantum lakes. Put more plainly, the hemidiabatic preparation of quantum lakes relies on the fact that we can sweep at a rate which suppresses transitions into high-energy states but allows transitions into low-energy states. The ansatz in Eq.~\eqref{eq:agp ansatz} is therefore already constructed to implement this ``cancel high, allow low'' procedure.
In other words, a poor approximation of the exact AGP, which targets the ground state, can be a great ``hemidiabatic gauge potential'' and target a quantum lakes state instead.
The second motivation of Eq.~\eqref{eq:agp ansatz} is that it contains nested commutators which appear naturally in the Magnus expansion \cite{Claeys_2019}. As such, even if the terms themselves seem unphysical, they can be realized with an appropriate Floquet protocol or pulse sequence.

Let us now consider the simplest example of the above approximate CD driving ideas: a qubit with the Landau-Zener Hamiltonian \cite{landau1932theorie,zener1932non,stuckelberg1932theorie,majorana1932atomi} 
\begin{align}
    H = -K Z - X .
    \label{eq:landau zener H}
\end{align}
For $K \to -\infty$, the ground state corresponds to $Z = -1$. As we sweep $K \to \infty$, the ground state swings around the Bloch sphere until it reaches the new ground state $Z = 1$. The rate at which we can perform such a sweep is set by the gap $\Delta = 2$ when $K = 0$. Physically, this process corresponds to implementing a $\pi$ pulse on the qubit, but the rate is limited by the timescale set by the gap.

If we instead calculate the first order AGP, we find $i [H, \partial_K H] \propto Y$, which is the generator of the $\pi$ pulse we implement during the sweep. It is easy to see that this is the exact AGP, since all other terms in the expansion return $Y$ as well. CD driving therefore applies a torque to our qubit and pushes it along during a fast sweep, ensuring it always follows the instantaneous ground state. Moreover, in the quench limit, the CD Hamiltonian reduces to just the driving term: $H_{\mathrm{CD}} \to \dot{K} \mathcal{A}_K$. We therefore see that a quench sweep corresponds to implementing a $\pi$ pulse in the usual sense: simply applying a magnetic field to rotate the qubit.

There are two key takeaways from this simple example. First, although the overall rate of any quantum process is set by the magnitude of the Hamiltonian, e.g. the laser power in a Rydberg atom array, evolution under the AGP uses this power more efficiently. 
While the adiabatic protocol must realize both the $X$ and $Z$ components of the field, only the latter sets the gap and restricts the preparation rate.
Evolving under the AGP focuses all power into the rotation, thereby accelerating the overall protocol even when laser power is constant. Second, the AGP can be interpreted as a force applied to specific degrees of freedom which, upon evolution for a finite time, can ``$\pi$ pulse'' the system from one state to another. We show how approximate AGPs constructed for preparing quantum lakes can be interpreted as forces on fast and slow defects and used to approximately ``$\pi$ pulse'' the system from a trivial state into an entangled quantum lakes state.

\section{Supplemental Materials B: Other Driving Protocols in the Qutrit}

\begin{figure}[t!]
	\centering
 
	\includegraphics[width=0.25\linewidth]{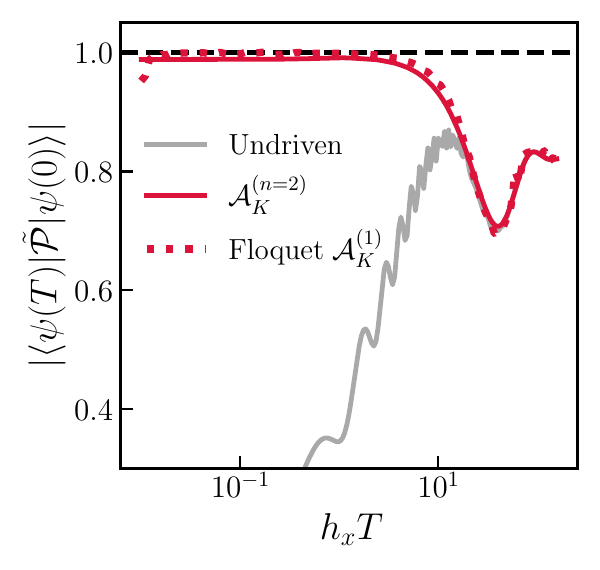}

	\caption{\textbf{Other AGPs in the Qutrit.} We plot the sweep fidelity after driving with the state specific AGP [Eq.~\eqref{eq:state specific agp}] and the Floquet implementation of $\mathcal{A}_K^{(1)}$ in the qutrit [Eq.~\eqref{eq:qutrit floquet protocol}]. The former cancels all transitions into and out of the second excited level while allowing all others and is therefore able to prepare the target projected state. The latter replicates the behavior from Fig.~\ref{fig:qutrit}(b) up to slight deviations.}
	\label{fig:other qutrit agps}
\end{figure}

In addition to the exactly gapped AGP from Eq.~\eqref{eq:exact gapped agp}, we also consider a state-specific AGP defined as
\begin{align}
    \bra{m} \mathcal{A}_{K}^{(n = 2)} \ket{n} \equiv (\delta_{n2} + \delta_{m2}) \bra{m} \mathcal{A}_{K} \ket{n} .
    \label{eq:state specific agp}
\end{align}
Similar to the gapped AGP, the state-specific AGP only cancels off transitions into and out of the second excited eigenstate of the Hamiltonian and allows all others.

We also consider a Floquet protocol (introduced in Ref.~\cite{Claeys_2019}) which stroboscopically generates the same dynamics as the Hamiltonian $H_{\mathrm{CD}} = H + \dot{K} \mathcal{A}_K^{(1)}$ from the text:
\begin{align}
    H_{\mathrm{F}} = \left( 1 + \frac{\omega}{\omega_0} \cos (\omega t) \right) H + \dot{K} \beta \sin (\omega t) \partial_K H  ,
    \label{eq:qutrit floquet protocol}
\end{align}
where $\omega = 10^4 h_x$ is the Floquet frequency, $\beta = 2 \omega_0 \alpha$ encodes the driving strength, and $\omega_0 = 10 h_x$ is a reference frequency that is much greater than $h_x$ but is otherwise arbitrary.
We use the same $\alpha$ from $\mathcal{A}_K^{(1)}$ in the text: 
\begin{align}
    \alpha = -1/(4 h_x^2 + h_z^2 + K^2) 
    \label{eq:qutrit alpha}
\end{align}
which follows from a straightforward minimization of the action in Eq.~\eqref{eq:trace action}.

Considering this protocol and $H_{\mathrm{CD}}^{(n=2)} \equiv H + \dot{K} \mathcal{A}_{K}^{(n=2)}$, we find the results in Fig.~\ref{fig:other qutrit agps}. 
These very nearly match those for $\Delta = 2 h_x$ in Fig.~\ref{fig:qutrit} for the same reason as before: large transitions (those which violate the $\mathcal{P} = 1$ condition) are canceled by the driving, while the small transitions that preserve the low-energy superposition are allowed. The reason for the slight difference in the state-specific case is that this method allows transitions to the first excited level for $K \ll -h_x$ while the gapped method does not. Evidently, this has a small effect on the final superposition in the low-energy subspace, but crucially, both methods support the ``cancel high, allow low'' strategy for designing AGPs.

The deviations in the Floquet procotol occur due to micromotion and higher order terms in the Magnus expansion.
We remark that this particular Floquet protocol could be experimentally infeasible as it requires a very large frequency $\omega$ and very strong driving amplitude ($\omega/\omega_0 = 10^3$). 
As we show in the Rydberg $\mathbb{Z}_2$ quantum spin liquid, protocols more amenable to direct experimental implementation can be easily designed.

\section{Supplemental Materials C: Additional Information for the Rydberg Ruby Lattice}

\subsection{$e$ and $m$ defects in the Rydberg ruby lattice}

We now provide more details on the $e$ and $m$ anyons present in the Rydberg model \cite{Verresen_2021,Sahay_2023} (along with the loop operators defined in the End Matter) and their dynamics during our sweeps.
Note that although we cannot ensure that the AGPs do not interact with the $m$ sector of the model, the driving inherently targets the quantum spin lakes regime. 
Indeed, in the initial state, the expectation value $\langle W_{p} \rangle = 0$ and must therefore increase to $\approx 1$ during the sweep if we wish to prepare an RVB-like state. One may ask how we can argue that $m$ dynamics are frozen if $\langle W_{p} \rangle$ evolves during the sweep. The reason is that the Wilson loop does not evolve if we restrict to the Gauss subspace where $G_v = -1$. Here, $\langle \widetilde{W}_p \rangle \equiv \langle \tilde{\mathcal{P}}_G W_p \tilde{\mathcal{P}}_G \rangle$ starts and remains near 1 for the entire sweep. As such, $\langle W_{p} \rangle$ increases due to the increasing population in this subspace, not because of some nontrivial $m$ dynamics.

\begin{figure}[t!]
	\centering

    \includegraphics[width=\linewidth]{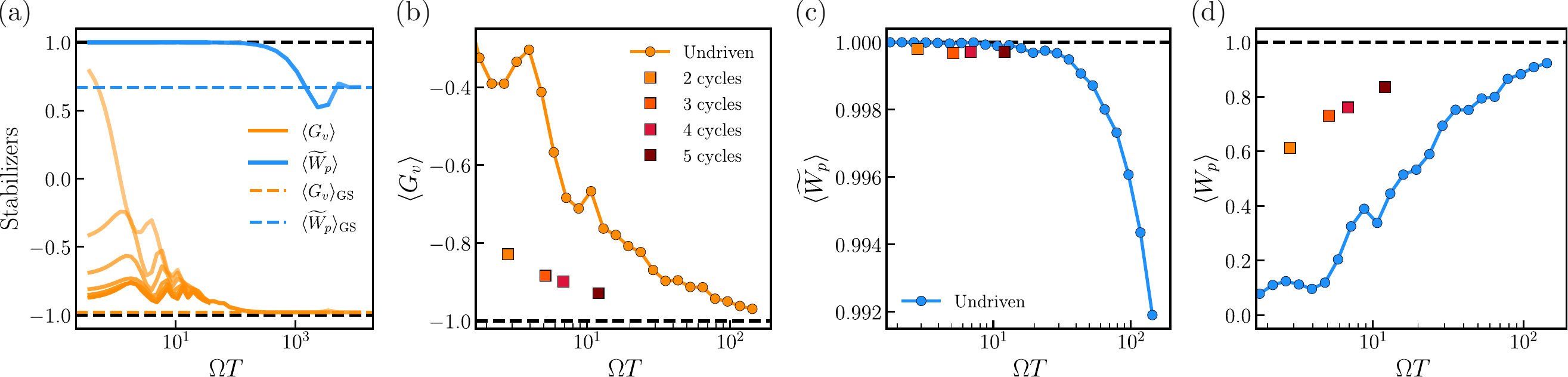}

	\caption{
    \textbf{Quasiparticle Observables in the Rydberg Ruby Lattice.}
    (a) We plot RVB stabilizer expectation values at the end of the sweeps from Fig.~\ref{fig:ruby acd driving} from undriven (lightest) to $\mathcal{A}_{\delta}^{(5)}$ (darkest). The clear separation of timescales allows for the hemidiabatic preparation of the spin lakes state, and approximate CD driving only targets the fast $e$ defects.
    (b) We also show the Gauss law $\langle G_v \rangle$ as a function of total preparation time for the undriven sweep in (a) and the pulse sequences in Fig.~\ref{fig:ruby speedup}. As the sweep rate is decreased, the Gauss law approaches $-1$, signifying the absence of $e$ excitations in the state.
    (c) To quantify the $m$ excitation dynamics within the low-energy subspace, we consider the Wilson loop of the state projected into the dimer covering subspace $\langle \widetilde{W}_p \rangle$. The values near 1 indicate the absence of $m$ excitations, which---coupled with the absence of $e$ excitations that would take the state out of this subspace---indicates that the states approximately realize the RVB state.
    (d) The full Wilson loop $\langle W_p \rangle$, in contrast, is much lower for quench sweeps and only increases as $\langle G_v \rangle \to -1$ and the population in the dimer covering subspace increases.
    This stabilizer value is used to calculate the lake size and quantify the spatial extent of the RVB order.
    }
	\label{fig:ruby stabs}
\end{figure}

This explanation of hemidiabaticity is supported by the numerics if we plot $\langle G_v \rangle$ and $\langle \widetilde{W}_p \rangle$ for the sweeps in Fig.~\ref{fig:ruby acd driving}.
These values are plotted in Fig.~\ref{fig:ruby stabs}(a). It is clear that the crossover from quench to hemidiabatic behavior is due to the $e$ timescale, and the crossover from hemidiabatic to adiabatic behavior is due to the $m$ timescale. Moreover, the approximate CD driving only targets the $e$ sector, forcing $\langle G_v \rangle \to -1$ without affecting the $m$ sector. 
In Figs.~\ref{fig:ruby stabs}(b-c), we plot these values for the optimized pulse sequences from Fig.~\ref{fig:ruby speedup}.
Just as with approximate CD driving, the pulse sequences reduce the density of $e$ excitations as $\langle G_v \rangle \to -1$ while leaving $\langle \widetilde{W}_p \rangle$ frozen near 1.

Finally, we see in Fig.~\ref{fig:ruby stabs}(d) that the full Wilson loop $\langle W_p \rangle$ is suppressed relative to $\langle \widetilde{W}_p \rangle$ due to the population outside the dimer covering subspace. Indeed, as this population is increased ($\langle G_v \rangle \to -1$), $\langle W_p \rangle$ follows the same trend as the Gauss law.
Although this value does not provide a good heuristic for dynamics within the low-energy subspace, it is necessary to calculate the size of the RVB lake.

\subsection{Approximate CD driving in the Rydberg ruby lattice}

For our numerics, we consider a $2 \times 3$ unit cell kagome lattice with periodic boundary conditions along its lattice vectors. We restrict to the translation and inversion symmetric subspace such that the 36 qubit space is reduced to a dimension of $11438 \equiv 2^{N_d}$.
The $\alpha_k$ in Eq.~\eqref{eq:ruby agp} are optimized according to the trace action in Eq.~\eqref{eq:trace action}, but the blockade projectors make analytic calculations of the trace difficult albeit possible. We instead use our numerically constructed $PXP$ and $PYP$ operators in the translation and inversion symmetric subspace to calculate the traces numerically for a $2\times2$ unit cell lattice. It has been shown that these optimized driving protocols generally do not scale with system size \cite{Sels_2017,Gjonbalaj_2022}, and we observe that the $2\times2$ traces give driving which prepares states efficiently on the larger $2\times3$ lattices in our numerics. As such, scaling this particular method to larger systems does not require more intensive optimization.

We also multiply the AGP by an overall factor $\lambda_f$ to maximize the final RVB fidelity of a quench sweep (we find $2.5 \lesssim \lambda_f \lesssim 3$) \cite{li2024quantumcounterdiabaticdrivinglocal}.
As such, the results in Fig.~\ref{fig:ruby acd driving} are simulated under the Hamiltonian $H + \dot{\delta} \lambda_f \mathcal{A}_{\delta}^{(\ell)}$.
One might worry that this optimization invalidates our argument that approximate AGPs \textit{naturally} target the hemidiabatic regime, but this is not the case.
Indeed, \textit{both} the RVB and true ground state overlap display maxima at nearly or exactly the same $\lambda_f$ values.
This is because the value of $\lambda_f \geq 1$ is accounting for the fact that $\delta \neq \infty$ at the end of the sweep, meaning that the prepared state is not the fixed-point RVB state; instead, it is dressed by perturbation theory and includes some virtual $e$ excitations.
As such, $\lambda_f \neq 1$ does not change how the driving interacts with $m$ excitations; it simply allows it to eliminate $e$'s more efficiently.
As another example of this effect, consider the Landau-Zener model in Eq.~\eqref{eq:landau zener H}. If one sweeps from a finite negative $K$ to a finite positive $K$, the qubit will never complete a full $\pi$ pulse and will end with $\langle Z \rangle < 1$.
To address this with CD driving, one can increase the strength of the driving under $Y$ such that the qubit is pushed slightly farther along its trajectory, achieving $Z = 1$ (the state analogous to the RVB state in the Rydberg model).

\begin{figure}[t!]
	\centering

    \includegraphics[width=0.75\linewidth]{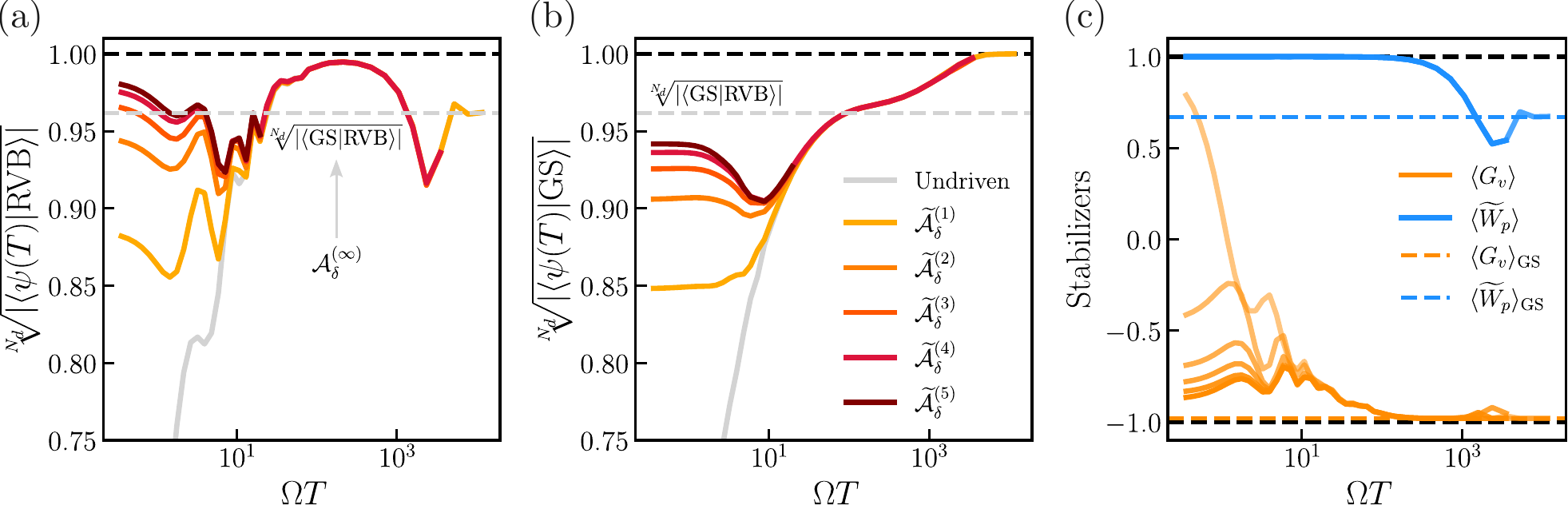}
	
	\caption{
    \textbf{Approximate CD Driving with a Simpler AGP Ansatz.}
    (a) We plot the fidelity of preparing the RVB state using the ansatz $\widetilde{\mathcal{A}}_{\delta}^{(\ell)}$ in Eq.~\eqref{eq:ruby PXP agp} for the same sweep as in Fig.~\ref{fig:ruby acd driving}. 
    For $\ell=1,2$, this ansatz is the same as the one used in the text, but at higher orders this ansatz leaves out some terms. 
    This simpler form of the driving allows us to simulate much longer sweeps at higher order.
    We find that the fidelity changes only marginally for most sweep rates, with the exception being $\Omega T \approx 10$ where we see a dip.
    (b) The same observations hold if we plot the overlap with the final VBS ground state instead.
    (c) The stabilizer expectation values for the sweep are also plotted. 
    Here, the dip becomes a slight increase in $\langle G_v \rangle$ near the same value of $\Omega T$.
    }
	\label{fig:ruby PXP ansatz}
\end{figure}

\subsection{Other driving protocols in the Rydberg ruby lattice}

We have argued that our results apply to a large class of approximate AGPs. 
We also consider pulse sequences which do not realize all terms in Eq.~\eqref{eq:ruby agp} but rather a subset of them.
As such, we now consider implementing approximate CD driving using the following restricted ansatz for the AGP:
\begin{align}
    \widetilde{\mathcal{A}}_{\delta}^{(\ell)}(\delta) =  \sum_{k=1}^{\ell} \widetilde{\alpha}_k(\delta) \underbrace{[PXP,\dots [PXP}_{2k-2},  P Y P ]] ,
    \label{eq:ruby PXP agp}
\end{align}
which uses only those terms that appear in the effective Hamiltonian of the pulse sequence in Eq.~\eqref{eq:ruby pulse eff H}.
For $\ell = 1,2$, the two forms are actually equivalent, but once we go to third order and beyond, new terms appear in the full ansatz.

The results of optimizing and driving with this ansatz, using the same procedure outlined above, are shown in Fig.~\ref{fig:ruby PXP ansatz}. As expected, the performance is not dramatically different when compared with that of the full ansatz. In the quench limit, there is very little change in the fidelities and stabilizer values. This supports the claim that the exact form of the AGP is not crucial to the hemidiabatic argument. It also provides some explanation for why the pulse sequence is still able to achieve a high preparation fidelity despite the missing terms in Eq.~\eqref{eq:ruby pulse eff H}.

Although we do not have a full explanation for the dip in fidelity visible near $\Omega T \approx 10$, such fluctuations are not uncommon in CD driving (see e.g. \cite{Sels_2017,Gjonbalaj_2022,Cepaite_2023}), and we suspect that it may arise from some kind of destructive interference between the AGP and Hamiltonian due to the missing AGP terms. We leave a full investigation to future work.

\subsection{Understanding the pulse sequence in the Rydberg ruby lattice}

\begin{figure}[t!]
	\centering
 
	\includegraphics[width=0.25\linewidth]{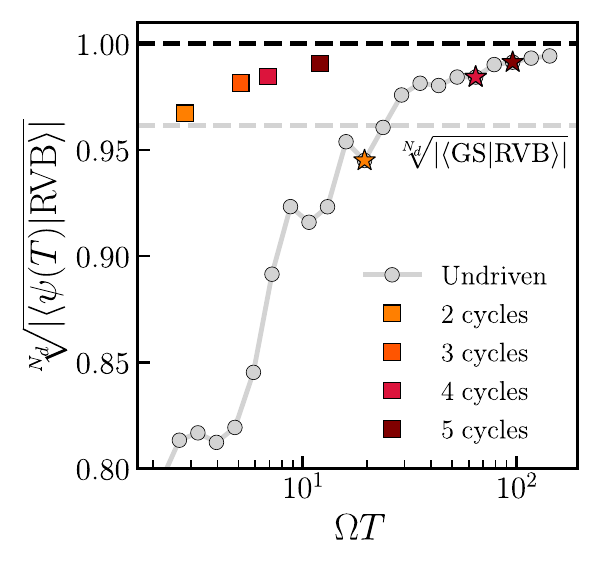}

	\caption{\textbf{Matching States Prepared by Pulse Sequences and Undriven Sweeps.} 
    To confirm that our pulse sequences generate spin lake states in the same fashion as a hemidiabatic sweep, we take the final state of each pulse sequence and find the undriven sweep for which the final overlap is maximized.
    These undriven sweeps are indicated by the colored stars (the 3 cycle and 4 cycle sequences share the same optimal $T$ value) along with the same data plotted in Fig.~\ref{fig:ruby speedup}(b).
    The overlaps between the states prepared by these sweeps and the corresponding pulse sequences are all $\sqrt[N_d]{|\langle \psi_{\mathrm{sweep}}(T) | \psi_c \rangle |} \approx 0.99$.
    This matching confirms the order of magnitude speedup observed in our other metrics.
    }
	\label{fig:ruby speedup match}
\end{figure}

Practically implementing shortcuts to adiabaticity often relies on simple sequences of laser pulses (see e.g. \cite{lukin2024quantumquenchdynamicsshortcut,Morawetz_2024}) which at first glance look very different from the continuous CD driving protocols analyzed above.
Here, we motivate the pulse sequence used in the main text from the driving terms in Eq.~\eqref{eq:ruby PXP agp}. The single cycle in Eq.~\eqref{eq:ruby cycle unitary} can be rewritten as
\begin{align}
    U_{c} =  \exp[-i y e^{-i x PXP} PYP e^{i x PXP}]  \exp[-i y e^{i x PXP} PYP e^{-i x PXP}] ,
\end{align}
where we have simply treated $e^{-i x PXP}$ as a basis rotation. The Baker-Campbell-Hausdorff expansion then tells us that the first dressed exponent can be expanded as
\begin{align}
    \exp[-i y e^{-i x PXP} PYP e^{i x PXP}] = \exp \left[-i y \sum_{k=1}^{\infty} \frac{(-ix)^{k-1}}{(k-1)!} \underbrace{[PXP,\dots [PXP}_{k-1},PYP ]] \right] ,
\end{align}
and similarly for the other dressed exponent but with $x \to -x$. Note that $k$ starts at 1 to agree with the convention in Eq.~\eqref{eq:ruby PXP agp}. Now taking the limit $y \ll 1$, we can find the effective Hamiltonian $H_c$ such that $U_c = \exp[- i H_c]$:
\begin{align}
    H_c = 2 y \sum_{k=1}^{\infty} \frac{(-ix)^{2k-2}}{(2k-2)!} \underbrace{[PXP,\dots [PXP}_{2k-2},PYP ]] + \mathcal{O}(y^2) .
\end{align}
Note that as long as $y$ is small enough to discard higher order Baker-Campbell-Hausdorff terms, the only terms present are those which also appear in Eq.~\eqref{eq:ruby PXP agp}. Rather than having a large set of parameters to optimize, however, we only have two. While $x$ encodes the magnitudes of $\widetilde{\alpha}_k$ relative to $\widetilde{\alpha}_1$, $y$ encodes the total time evolved under the AGP. 

Although this expansion clearly realizes the same terms as in Eq.~\eqref{eq:ruby PXP agp}, one might ask whether the sequence $(-ix)^{2k-2}/(2k-2)!$ is a good approximation for the optimal coefficients $\widetilde{\alpha}_k$ found by the trace action.
Indeed, the two degrees of freedom $x$ and $y$ are enough to ensure that $\widetilde{\alpha}_1$ and $\widetilde{\alpha}_2$ are exactly reproduced, but the higher order terms may or may not be matched by the sequence.
Empirically, we find that the sequence is often able to approximate $\widetilde{\alpha}_3$, $\widetilde{\alpha}_4$, and $\widetilde{\alpha}_5$ to within an order of magnitude even when $x$ and $y$ are chosen only using $\widetilde{\alpha}_1$ and $\widetilde{\alpha}_2$.
However, this difference is still too large to reproduce the behavior found in Fig.~\ref{fig:ruby PXP ansatz} using the pulse sequence.
In fact, this sensitivity of the dynamics to changes in the driving has been observed before in CD driving (see e.g. \cite{Gjonbalaj_2022}).
Thus, we instead optimize $x$ and $y$ such that the higher order terms can still contribute despite not reproducing the $\widetilde{\alpha}_k$.

Although the fidelity densities, stabilizer values, and lake sizes all confirm that these optimized pulse sequences speed up lake preparation by about an order of magnitude, we can make this claim more precise by asking which undriven sweep prepares a state $\ket{\psi_{\mathrm{sweep}}(T)}$ closest to that prepared by a given pulse sequence $\ket{\psi_c}$. 
This data is plotted in Fig.~\ref{fig:ruby speedup match} with the same data from Fig.~\ref{fig:ruby speedup}(b). 
For each pulse sequence, we find the value of $T$ which maximizes the quantity $\sqrt[N_d]{|\langle \psi_{\mathrm{sweep}}(T) | \psi_c \rangle |}$ and mark this value with a colored star.
We consistently find a maximal overlap of $\sqrt[N_d]{|\langle \psi_{\mathrm{sweep}}(T) | \psi_c \rangle |} \approx 0.99$, confirming that the pulse sequences generate spin lakes similar to those generated by hemidiabatic sweeps.
Moreover, the order of magnitude speedup in RVB preparation is clearly visible when comparing stars and squares of the same color.

\subsection{System size independence of the results}

\begin{figure}[t!]
	\centering

    \includegraphics[width=\linewidth]{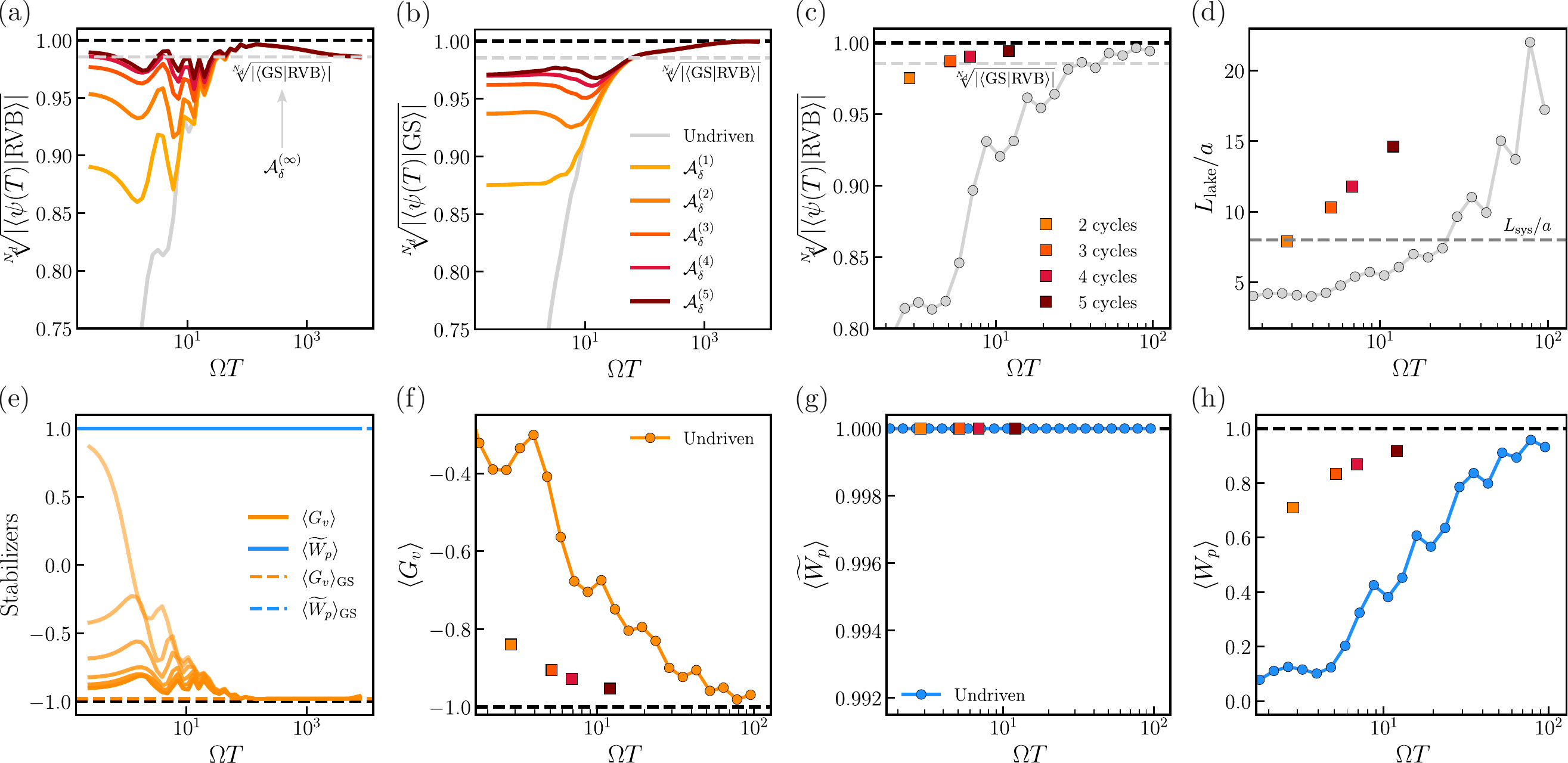}
	
	\caption{
    \textbf{Driving in a 24 Qubit Ruby Lattice.}
    We use the same approximate CD driving protocols from Fig.~\ref{fig:ruby acd driving} and the same pulse sequences from Fig.~\ref{fig:ruby speedup} and apply them in a 24 qubit Rydberg ruby lattice. 
    (a) Although the overlap density of the RVB state with the true ground state is nearly unity (due to the absence of a VBS phase at this system size), the quench values of each sweep are only marginally higher for 24 qubits than 36 qubits.
    (b) The absence of the VBS phase means that the quench values of ground state overlap are higher than in the larger system, but it is clear that the driving still targets the hemidiabatic regime instead of the adiabatic regime.
    (c) The same pulse sequences still lead to about an order of magnitude speedup in RVB state preparation despite the changes in the fidelity density.
    (d) Similarly, the sizes of the lakes prepared by these sequences are slightly larger in this smaller system, but the speedup remains about the same.
    (e) As expected, the Gauss law values for the approximate CD driving simulations are slightly closer to $-1$ than in the 36 qubit system.
    However, the absence of the VBS phase means that the Wilson loop $\widetilde{W}_p$ in the dimer covering subspace does not drop in the adiabatic limit.
    This is not a cause for concern as we have already shown that the AGP only has a significant effect for sweeps faster than hemidiabatic, where the $m$ excitations are frozen anyway.
    (f) As expected, the Gauss law values of states prepared by the pulse sequences do not change much with system size.
    (g) In addition, the sequences preserve the complete absence of $m$ excitations in the dimer covering subspace, keeping $\widetilde{W}_p = 1$.
    (h) Finally, the expectation values of the full Wilson loop $\langle W_p \rangle$ are marginally larger than those for 36 qubits but still track the population in the dimer covering subspace.
    }
	\label{fig:ruby 2x2 all}
\end{figure}

Numerical evidence suggests (see e.g. \cite{Gjonbalaj_2022}) that the optimized $\alpha_k$ do not scale with system size. Since the AGP acts locally to create the desired order, this makes physical sense. To argue that our driving protocols do not scale with system size, we provide two forms of evidence. First, we will now show that the same approximate CD driving and optimized pulse sequences give similar (although slightly better) performance in a smaller Rydberg ruby lattice of 24 qubits.
In Supplemental Materials E,
we perform matrix product state (MPS) numerics for a modified toric code model on an infinite cylinder, showing that many of our findings extend to the thermodynamic limit.

By using the same approximate CD driving protocol from the text in a $2\times2$ unit cell Rydberg ruby lattice, we find the results in Fig.~\ref{fig:ruby 2x2 all}(a-b). Compared to the results for the $2\times3$ lattice, the RVB fidelities in the quench limit are marginally higher.
If we now take the $x,y$ values used in Fig.~\ref{fig:ruby speedup} and apply them in the 24 qubit system, we find the results in Fig.~\ref{fig:ruby 2x2 all}(c-d).
Aside from some small shifts, the fidelity densities do not dramatically change, and although the lake sizes are a few lattice spacings larger than those found in the $2\times3$ system, the same speedup of nearly an order of magnitude is observed.

To further investigate the driving performance in this system, we calculate the same stabilizers in Fig.~\ref{fig:ruby 2x2 all}(e-h) for the $2\times2$ lattice as in Fig.~\ref{fig:ruby stabs}.
Although the values of $\langle G_v \rangle$ and $\langle W_p \rangle$ are marginally better, there is a glaring qualitative difference in the behavior of $\langle \widetilde{W}_p \rangle$. In particular, this value does not decrease from 1 as the sweep becomes more adiabatic.
This is because the final ground state actually has $\langle \widetilde{W}_p \rangle = 1$ for the smaller system. The VBS ``phase'' only appears once the system size reaches $2\times3$ unit cells, a consequence of the fact that for $2\times2$ unit cells, one cannot move from one dimer covering to another by applying the Wilson loop around a single plaquette.
As a result, the ground state has no $e$'s or $m$'s, but it is not exactly the RVB state. Rather, it is a different ``logical'' state of the system, related to the RVB state by the application of a string operator which wraps around the torus. This accounts for the overlap between the ground and RVB states not reaching unity but still being higher than in the larger system.
Although one might worry that this large qualitative difference between the system sizes invalidates any argument we make about system size independence, it is clear that in the quench limit, the $m$'s cannot affect the dynamics because they are frozen. As such, whether or not they appear in the ground state is immaterial; rather, one can see that the $e$ dynamics are largely the same, as we would expect.

Another concern which must be addressed is whether the absence of the VBS phase artificially biases our optimized driving parameters (which are calculated using operator traces in the $2\times2$ system) to target the RVB state.
To see that this is not a concern, consider the driving parameter $\alpha$ from Eq.~\eqref{eq:qutrit alpha} used in the qutrit.
The small splitting of the low-energy subspace in the qutrit is controlled by $h_z \ll h_x$, and this separation of energy scales defines the hemidiabatic regime.
It is clear that setting $h_z \to 0$ in Eq.~\eqref{eq:qutrit alpha} barely changes the driving protocol as it is already such a small contribution to $\alpha$, even though this change drastically affects the ground state for large $K$, making it the target state $\frac{1}{\sqrt{2}} (\ket{1} + \ket{-1})$.
Indeed, this same structure is present in the Landau-Ginzburg model considered above and the deformed toric code model considered in Supplemental Materials E.
More generally, the existence of the hemidiabatic regime means that ignoring the small energy scales of the slow excitation does not substantially change the driving.
Similarly, the perturbative splitting of the dimer coverings in the 36 qubit ruby lattice is too small compared to the detuning $\delta$ to substantially modify the $\alpha$'s.
Moreover, note that the values of $\langle \widetilde{W}_p \rangle$ are \textit{completely} unaffected by the driving in Fig.~\ref{fig:ruby stabs}(a) and Fig.~\ref{fig:ruby PXP ansatz}(c).
This is not due to some fine tuning of the driving parameters $\alpha_k$ but rather due to the inability of the approximate AGP to drive $m$ excitation dynamics at all.

These small changes in the driving performance as a function of the number of qubits therefore provide preliminary support for the claim that our results do not scale with system size.
Later in the SM, we will show that MPS simulations also give similar results in the thermodynamic limit.

\section{Supplemental Materials D: Approximate CD Driving in a Landau-Ginzburg Model}

In this section, we will review the statements made in the Semi-Classical Picture section of the main text and provide further explanation for how hemidiabatic preparation is able to target the lakes state.

We consider the simple two-component Landau-Ginzburg field theory with the potential 
\begin{align}
    V = \frac{1}{2} K \phi_a^2 + \frac{\lambda_a}{4!} \phi_a^4 + \frac{1}{2} (h_x \phi_a^2 - h_z) \phi_b^2 + \frac{\lambda_b}{4!} \phi_b^4 ,
\end{align}
describing a gas of two bosons $a$ and $b$.
For $K$ large and negative, the $a$ bosons form a Bose-Einstein condensate where $\langle \phi_a \rangle \neq 0$. 
As $K$ becomes large and positive, the ground state transitions to a condensate of $b$ bosons, such that $\langle \phi_b \rangle \neq 0$.
These bosons can be interpreted as defects in an otherwise ordered phase; as such, when they condense, the order is destroyed.
It is clear that the ordered phase, where no condensate is present, is not accessible using adiabatic sweeps of the parameter $K$, which always keeps the system in equilibrium.

By simply separating the intrinsic energy and timescales of $a$ and $b$, we can realize a hemidiabatic regime and prepare an ``ordered'' state with no defects.
To implement dynamics, the full Hamiltonian reads
\begin{align}
    H =  \int d^2 x \left[ \sum_{i \in \{a,b\}}\frac{1}{2} ( \Delta_i \Pi_i^2 + f_i (\nabla \phi_i)^2)  + V  \right] ,
\end{align}
where $\{\Delta_i\}$ control the timescales of each boson and $\{f_i\}$ control the spatial couplings. 
For the truncated Wigner approximation simulations shown in Fig.~\ref{fig:full LG numerics}, we use the following parameters: $\Delta_a = 0.05h_x$, $\Delta_b = 0.001 h_z$, $f_a = 0.1 h_x$, $f_b = 0.01 h_x$, $h_z = h_x/5$, $\lambda_a = 100 h_x$, and $\lambda_b = 0.05 h_x$. We use a 2D torus of $10 \times 10$ anharmonic oscillators to discretize the space of our field theory. The amount of quantum fluctuations is controlled by $\hbar$, which we set to 1.

To simplify sampling from the initial many-body ground state's Wigner function, we first compute the ground state wavefunction of an uncoupled site. This state is nearly a product of Gaussians in both $\phi_a$ and $\phi_b$ with $>0.9989$ fidelity, so we approximate it as such. This makes the associated Wigner function a product of Gaussians as well, simplifying the sampling procedure. For each site, we sample $\phi_a$, $\Pi_a$, $\phi_b$, and $\Pi_b$ independently from these Gaussians and then classically evolve the entire system while adiabatically sweeping $f_a,f_b$ linearly from 0 to their final values. We are therefore able to account for spatial fluctuations and correlations without calculating the entire system's Wigner function. The values for $f_a,f_b$ are chosen such that the condensed phases in the mean field treatment are not washed out by spatial fluctuations. We randomly initialize 10 million system states in this way to include quantum fluctuations in our classical dynamics.

Each initialization is classically evolved while $K$ is swept linearly from $-20h_x$ to $20h_x$ in a time $T$. Although the order parameters $\langle \phi_a \rangle$ and $\langle \phi_b \rangle$ (where $\langle \rangle$ indicates averaging over initializations) are the most intuitive probes of whether each boson has condensed, these quantities can still be zero if the classical trajectories choose different degenerate ground states in the double well potential. In other words, the overall Wigner distribution will still be $\mathbb{Z}_2$ symmetric because the dynamics preserve the symmetry. To address this, we consider the $\mathbb{Z}_2$ off-diagonal long range order parameters $\langle a_{i}^{\dagger}(\mathbf{r}) a_i(\mathbf{0}) \rangle$ in analogy with the superconducting case, where 
\begin{align} 
    a_a(\mathbf{r}) &= \left( \frac{K_f}{4 \Delta_a} \right)^{1/4} \left( \phi_a(\mathbf{r}) + \frac{i}{\sqrt{K_f/\Delta_a}} \Pi_a (\mathbf{r}) \right) , \nonumber \\
    a_b(\mathbf{r}) &= \left( \frac{h_z}{4 \Delta_b} \right)^{1/4} \left( \phi_b(\mathbf{r}) + \frac{i}{\sqrt{h_z/\Delta_b}} \Pi_b (\mathbf{r}) \right) ,
\end{align}
and $K_f = 20h_x$ is the final value of $K$. 
One can show (see e.g. Appendix F of \cite{sahay2024superconductivity}) that this encodes the same information as the order parameters for $\mathbf{r} \to \infty$. 
The parameters are chosen using the on-site harmonic terms of the Hamiltonian at the end of the sweep and mapping them onto harmonic oscillators in each defect sector.
In our simulations, we choose $\mathbf{r} = (5,0)$
far from the point $\mathbf{0} = (0,0)$ on our torus while still keeping the signal to noise ratio reasonable. To eliminate spatial fluctuations and enforce translational invariance, we compute the average of this quantity over the lattice:
\begin{align}
    \langle \langle a_i^{\dagger}(\mathbf{r}) a_i(\mathbf{0}) \rangle \rangle \equiv 
    \left\langle \underset{\mathbf{R} \in \Lambda}{\mathbb{E}} a_i^{\dagger}(\mathbf{R} + \mathbf{r}) a_i(\mathbf{R}) \right\rangle ,
\end{align}
where $\Lambda$ is our set of lattice coordinates.

The optimization of AGPs in systems of oscillators, where the local phase space (or Hilbert space) is infinite dimensional, is not as straightforward as in systems of spins and fermions. Instead of using the infinite temperature trace action described in \cite{Sels_2017}, we use the method developed in \cite{Gjonbalaj_2022} of tracking the adiabatic evolution of a single distribution (or Wigner function) and then using this to optimize the driving. In addition to this optimization, we multiply the driving term by an overall factor $\lambda_f = 45 h_x$ chosen to minimize the $a$ order parameter $| \langle \langle a_a^{\dagger}(\mathbf{r}) a_a(\mathbf{0}) \rangle \rangle |$ in the quench limit with driving \cite{li2024quantumcounterdiabaticdrivinglocal}. 
As explained in the section on the ruby lattice, this does not invalidate our argument that approximate CD driving naturally targets the hemidiabatic regime because it is simply driving out more $a$ bosons and does not bias the $b$ boson dynamics in any way.

\section{Supplemental Materials E: Approximate CD Driving in the Deformed Toric Code}
\label{sm:dtc}

As a final example, we analyze how approximate CD driving can create quantum spin lakes in the deformed toric code (DTC) model of Ref.~\cite{Sahay_2023}. First, we briefly review the model and discuss the theory and analytics behind approximate CD driving. Then, we show numerical evidence for accelerating the preparation of quantum spin lakes.
As these numerics are performed for using matrix product states (MPS) on an infinite cylinder, it provides evidence that lakes preparation is not a finite-size effect. 
For a more detailed introduction to the model, see Ref.~\cite{Sahay_2023}.

\subsection{DTC model}

The DTC Hamiltonian reads
\begin{align}
    H_{\mathrm{DTC}} = -K \sum_v \starZs - h_x \sum_i X_i - h_z \sum_i Z_i ,
\end{align}
where $X_i,Y_i,Z_i$ are the spin-1/2 Pauli matrices for the qubit living on the $i$th link of the square lattice. We will only consider $K,h_x,h_z\geq0$.

For $h_z = 0$, this model has two simple fixed points. When $K = 0$, the ground state is simply the product state $\ket{+}^{\otimes N}$. When $K \to \infty$, the ground state is the same as the toric code \cite{Kitaev_2003} and is stabilized by a perturbative resonance term of strength $J_{\mathrm{eff}} \sim \mathcal{O}(h_x^4/K^3)$ that looks like
\begin{align}
    W_p = \loopXs .
    \label{eq:Wp}
\end{align}
Finally, as we increase $h_z$ from zero, a third phase emerges, the fixed point of which corresponds to $K,h_x = 0$ and the ground state $\ket{0}^{\otimes N}$. For the phase diagram of this model, see Ref.~\cite{Sahay_2023}.

At the $K \to \infty$, $h_x,h_z = 0$ fixed point, we can understand excitations in terms of the well-known $e$ and $m$ anyons of the original toric code. $e$ anyons are defined using the Gauss law term:
\begin{align}
    G_v = \starZs \: .
\end{align}
When $G_v = -1$, an $e$ anyon lives at vertex $v$. Similarly, when the Wilson loop operator $W_p = -1$, an $m$ anyon lives on plaquette $p$.
When both equal 1, there are no anyons present, and the system is in one of the logical states of the toric code.
The nearby phases can then be understood in terms of these anyons.
As $h_x/K$ increases away from the ``deconfined'' toric code phase, virtual $e$ anyons start to appear in the ground state. Once $h_x/K$ becomes large enough, these anyons condense in the ground state and drive a phase transition into the ``Higgs'' phase, which is adiabatically connected to the $K, h_z = 0$ fixed point. Similarly, as $h_z$ increases relative to the perturbative $m$ anyon gap $J_{\mathrm{eff}} \sim \mathcal{O}(h_x^4/K^3)$, virtual $m$ anyons appear in the ground state before finally condensing and driving a transition into the ``confined'' phase, which is adiabatically connected to the $K,h_x = 0$ fixed point.

It was shown in Ref.~\cite{Sahay_2023} that initializing this system in the ground state with $K = 0$ and $h_z / h_x \leq \sqrt{\frac{\sqrt{2}-1}{2}} \approx 0.46$ and applying the Gauss law projector
\begin{align}
    \mathcal{P}_G = \prod_v \frac{1 + G_v}{2}
\end{align}
will result in a quantum spin liquid. Dynamically, this can be implemented approximately to create quantum spin lakes if $K$ is increased from 0 at a hemidiabatic rate. Crucially, we do not need to target the toric code phase of the model during this sweep; even if the final value of $K$ lands us in the confined phase, the fact that $h_z,J_{\mathrm{eff}}$ are small compared to $h_x,K$ means that $m$ anyons are inherently slower than $e$ anyons and will not have time to condense into the ground state.

Let us now consider how approximate CD driving will proceed in this model. The first order AGP is
\begin{align}
    \mathcal{A}_K^{(1)} = i \alpha \left[ H, \partial_K H \right] = 2 h_x \alpha \sum_{v,\mu} \starYZs ,
    \label{eq:dtc agp ansatz}
\end{align}
where $\mu \in \{0,1,2,3\}$ labels which leg of the vertex hosts the $Y$ operator. We will refer to the operator inside the sum as the ``starY'' operator. Before optimizing and driving with this AGP, let us consider its physical interpretation. As it is a multiplication of the Gauss operator and a local $X$ flip, the operator first detects the presence or absence of an $e$ anyon and, depending on the outcome, applies $\pm X$ to one bond of the vertex. Since this operator can nucleate pairs of $e$ anyons, or equivalently hop $e$ anyons across a bond, this operator applies a phase-dependent flip of the Gauss law on a given vertex, exactly analogous to the interpretation given in the Landau-Ginzburg treatment from the main text.
Finally, note that starY commutes with the $W_p$ stabilizers in Eq.~\eqref{eq:Wp}. As such, dynamics generated by this AGP cannot change the density of $m$ anyons (calculated as $\langle W_p \rangle$). As shown below, this fact provides a very simple method for generating higher-order AGPs that exactly commute with $W_p$.

\subsection{Optimization at first order}
\label{app:dtc agp opt}

We want to calculate the optimal $\alpha$ for the AGP in Eq.~\eqref{eq:dtc agp ansatz}. First, we have the $G_K$ operator:
\begin{align}
    G_K &= \partial_K H + i [\mathcal{A}_K^{(1)},H] \nonumber \\
    &= (-1 - 16 h_x^2 \alpha) \sum_v \starZs + 8 h_x K \alpha \left( \sum_i \starXstarZs + X_i \right) + 8 h_x^2 \alpha \sum_{v,\mu>\nu} \starYYZs + 4 h_x h_z \alpha \sum_{v,\mu} \starXZs .
\end{align}
In the first $8 h_x K \alpha$ term, $i$ labels the link where the $X$ lives. In the $8h_x^2 \alpha$ term, $\mu,\nu \in \{0,1,2,3\}$ indicate the positions of the two $Y$ operators but $\mu >\nu$ to avoid double counting. 

Because Pauli matrices are traceless and square to the identity, calculating the action is straightforward:
\begin{align}
    \mathcal{S} = \Tr[G_K^2] = \mathcal{D} N_v \Big( (1 + 16 h_x^2 \alpha)^2 + 2 (8 h_x K \alpha)^2 + 2 (8 h_x K \alpha)^2 + 6 (8 h_x^2 \alpha)^2 + 4 (4 h_x h_z \alpha)^2 \Big) ,
\end{align}
where $N_v$ is the number of vertices in the system and $\mathcal{D} = 2^{2 N_v}$ is the total dimension of the Hilbert space. Optimizing with $\partial_{\alpha} \mathcal{S} = 0$ yields
\begin{align}
    \alpha = - \frac{1}{4(10 h_x^2 + h_z^2 + 4 K^2)} .
    \label{eq:dtc optimized alpha}
\end{align}
Note that because we require $h_z/h_x \lesssim 0.46$, the $h_z^2$ term barely contributes to this protocol. Indeed, we might as well eliminate it all together:
\begin{align}
    \alpha = - \frac{1}{4(10 h_x^2 + 4 K^2)} .
    \label{eq:dtc optimized alpha hz=0}
\end{align}
This is the protocol we would have in a DTC model where $h_z = 0$. Conceptually this makes more sense: our AGP has no interaction with the $m$ anyon sector of the Hilbert space, so we should optimize it in a model where virtual $m$ anyons cannot affect the optimization. Indeed, when $h_z = 0$, $m$ anyons can never nucleate in the ground state, and no confined phase exists. We should therefore interpret this AGP as a force which only acts on $e$ anyons and drives them into and out of our state.
Although this change has essentially no effect at first order, it provides the road map for designing higher-order AGP ansatzes which do not interact with the $m$ sector.

Note that our optimization can be improved using an overall factor $\lambda_f$ as mentioned above. Because AGP evolution corresponds to using approximate CD driving in a quench sweep, the evolution and final observables are the same when the total time $h_x t$ under AGP evolution equals the total integrated driving strength in approximate CD driving:
\begin{align}
    h_x t = -\int_{0}^{4 h_x} \alpha dK = \frac{\arctan \left( 4 \sqrt{\frac{2}{5}} \right)}{8\sqrt{10}} \approx 0.047 ,
\end{align}
using Eq.~\eqref{eq:dtc optimized alpha hz=0}. 
However, to truly reach the fixed-point toric code state, we need to sweep to $K \to \infty$. This modifies the integral to give $\pi/(16 \sqrt{10}) \approx 0.062$.
Although we cannot sweep to infinite $K$ in experiment, we can modify the driving strength to ``complete the $\pi$ pulse'' while doing a sweep to finite $K$  
\cite{li2024quantumcounterdiabaticdrivinglocal} using
\begin{align}
    \lambda_f = \frac{\pi}{2 \arctan \left( 4 \sqrt{\frac{2}{5}} \right)} \approx 1.315 .
    \label{eq:dtc lambdaf}
\end{align}
Although this method is not exact, and the variational procedure is not guaranteed to maximize the QSL fidelity (since it is optimized for all states in the spectrum), all calculations are analytic and display quantum spin lake signatures in the thermodynamic limit as shown below.

\subsection{Approximate CD driving}

\begin{figure}[t!]
	\centering

    \includegraphics[width=\linewidth]{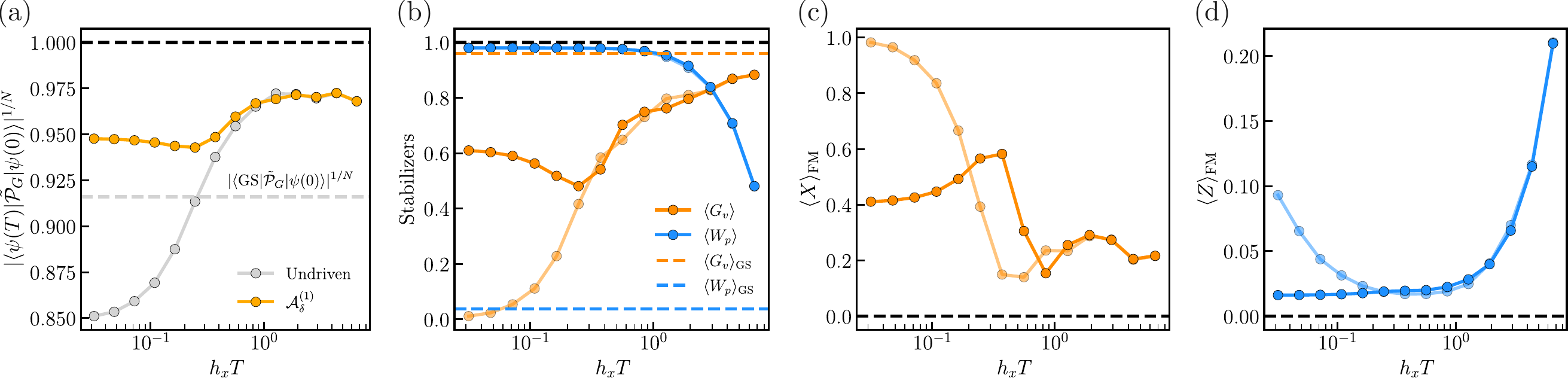}

	\caption{\textbf{Approximate CD Driving in the Deformed Toric Code.} 
    We implement approximate CD driving while ramping $K \in [0,4h_x]$ on an infinite cylinder with $L_y = 2$ and $h_z = 0.1 h_x$. 
    We use a bond dimension of $\chi = 64$ and time step of $h_x dt = 0.005\min(1, h_x T)$, except for the largest value of $h_x T$, which uses $\chi = 128$ and $h_x dt = 0.001$.
    (a) By plotting the local overlap of the final state and the target state as a function of total sweep time, we recover the same hemidiabatic behavior seen in our previous examples.
    (b) In addition, the separation of timescales between the $e$ and $m$ anyons is evident by plotting the stabilizer expectation values as a function of total sweep time.
    Just as in the Rydberg case, the driving only affects $e$ anyons during fast sweeps and does not affect the dynamics substantially otherwise.
    We also observe a small dip in performance around $h_x T \approx 0.3$ similar to the dip found in Fig.~\ref{fig:ruby PXP ansatz}.
    (c) We show the $X$ FM order parameter to quantify the condensation of $e$ anyons and see that the driving decreases the value in line with our previous results.
    Because the cylinder has circumference $L_y = 2$, we measure these order parameters on a loop around a single plaquette (see Fig.~4 of Ref.~\cite{Sahay_2023}).
    (d) Similarly, the $Z$ FM order parameter (which quantifies the condensation of $m$ anyons) is mostly unchanged except for a decrease for quench sweeps.
    Now, we measure around a single dual plaquette such that the loop surrounds a single vertex.
    }
	\label{fig:dtc acd driving}
\end{figure}

Using the result from Eq.~\eqref{eq:dtc lambdaf}, we can implement approximate CD driving under the Hamiltonian
\begin{align}
    H_{\mathrm{ACD}} = H_{\mathrm{DTC}} + \dot{K} \lambda_f \mathcal{A}_K^{(1)} .
\end{align}
The results of evolving under this Hamiltonian while sweeping $K$ from $0$ to $4 h_x$ in the confined phase over a time $h_x T$ are shown in Fig.~\ref{fig:dtc acd driving}. Tensor network simulations were performed using the TeNPy library \cite{tenpy,hauschild2024tensornetworkpythontenpy} on an infinite cylinder using the same construction as Ref.~\cite{Sahay_2023}. Due to the complexity of the driving, we were only able to reach a circumference of $L_y = 2$ with our numerics, but the infinite system size in one of the dimensions (in combination with the data from the ruby lattice simulations) shows the effectiveness of our methods in the thermodynamic limit.

In Fig.~\ref{fig:dtc acd driving}(a), we plot the fidelity per qubit of being in the projected state as a function of the total sweep time. The undriven ramp, as shown in \cite{Sahay_2023}, increases from a small value for quench sweeps up to a spin lakes peak around $h_x T = 1$ before eventually sinking down to the adiabatic fidelity in gray, given by the overlap between the final ground state and the target projected state. As expected, the first order AGP does not target the ground state but rather the spin lakes state prepared at intermediate sweep rates. In the DTC, this has a very clear explanation: the AGP only affects $e$ anyon dynamics and therefore explicitly cannot push the system into the confined phase on its own. All $m$ anyon nucleation is controlled by $H_{\mathrm{DTC}}$ dynamics. 

This is clearly seen in Fig.~\ref{fig:dtc acd driving}(b), where we plot the stabilizer values for both the undriven and driven sweeps as a function of total sweep time. While the Wilson loops are unaffected (up to numerical error), the Gauss law is systematically increased near the quench limit as an (approximate) extension of the spin lakes peak. The slight differences in the Gauss law  between the driven and undriven sweeps occur because of oscillations in these stabilizers during the tail end of the sweep. Each sweep 
ends at a slightly different point in the oscillation period, giving slight disagreements between the two curves.
Aside from these dips, it is clear that the driving increases the Gauss law and improves the overlap with the QSL state. 

Finally, to probe the long-range correlations of the state, we also consider Fredenhagen-Marcu (FM) string order parameters \cite{Verresen_2021,Chandran_2013,Fredenhagen_1983,Fredenhagen_1986,1986_MARCU,Gregor_2011}, defined in \cite{Sahay_2023} and plotted in Fig.~\ref{fig:dtc acd driving}(c-d).
Intuitively, these parameters probe the condensation of anyons similar to the ODLR order parameters used in the Landau-Ginzburg model. If both string order parameters flow to zero with increasing string length, the state corresponds to the deconfined spin liquid phase where no anyons have condensed.
We see that the overall effect of the driving is to reduce the FM order parameters for fast sweeps. The one exception to this is the peak in $\langle X \rangle_{\mathrm{FM}}$ near $h_x T \sim 0.3$ which lines up with the dip in the Gauss law in Fig.~\ref{fig:dtc acd driving}(b). The exact reason for this is unclear and deserves further attention, especially because it also appears in our results in Fig.~\ref{fig:ruby PXP ansatz}.

\subsection{Higher order AGPs}

Having numerically shown that the first order AGP effectively targets the hemidiabatic regime where quantum spin lakes are prepared, our next question is naturally how to extend this effect to higher orders. While we argue that the unmodified variational procedure from \cite{Sels_2017,Claeys_2019} already inherently targets this regime, the DTC allows us to construct AGPs which \textit{exactly} conserve $W_p$.
Note that this is fundamentally different from our use of the $2\times2$ ruby lattice (where no VBS phase exists) to optimize the $\alpha$ values used to drive the $2\times3$ ruby lattice.
In the larger system, the AGP is not constructed to conserve $\widetilde{W}_p$ but does so due to a separation of timescales, and will eventually reduce $\widetilde{W}_p$ at high enough order.
Here, we modify the AGP such that it commutes with $W_p$ at \textit{all} orders.

In particular, the clear separation of the $e$ and $m$ anyon sectors in the Hilbert space allows us to create a ``hemidiabatic gauge potential'' which commutes with $W_p$ at all orders. We modify Eq.~\eqref{eq:agp ansatz} to be
\begin{align}
    \mathcal{A}_K^{(\ell)} = i \sum_{k = 1}^{\ell} \alpha_k \underbrace{[H_e,[H_e,\dots[H_e}_{2k-1},\partial_K H_e]]] ,
    \label{eq:dtc hgp ansatz}
\end{align}
using
\begin{align}
    H_e = -K \sum_v \starZs - h_x \sum_i X_i ,
\end{align}
where we have simply set $h_z = 0$ in $H_{\mathrm{DTC}}$. This Hamiltonian commutes with all $W_p$, so we restrict the Hilbert space to the totally symmetric sector with $W_p = 1$ for all plaquettes. The Hamiltonian can therefore never nucleate $m$ anyons and has no confined phase. All AGPs generated using \eqref{eq:dtc hgp ansatz} therefore commute with $W_p$. More intuitively, we are constructing our AGPs in a variation of the DTC where $m$ anyons do not appear and importing these operators to target quantum spin lakes in a model which does feature $m$ anyons. At first order, Eq.~\eqref{eq:dtc hgp ansatz} reproduces the same protocol as in Eq.~\eqref{eq:dtc optimized alpha hz=0}. At second order, however, the difference becomes more apparent. While the original AGP would not commute with the plaquettes due to virtual $m$ anyon fluctuations, these terms would be small since $h_z \ll h_x$. When using $H_e$, these terms are automatically excluded, and the ansatz becomes
\begin{align}
    \mathcal{A}_K^{(2)} &= 2 h_x \alpha_1 \sum_{v,\mu} \starYZs + \alpha_2 \Bigg[ (80  h_x^3 + 32  h_x K^2) \sum_{v,\mu} \starYZs \nonumber \\
    &+ (-48 h_x^3) \sum_{v,\mu} \starYYYZ + (-32 h_x^2 K) \sum_{\substack{i ,\mu}} \starXstarYZs \Bigg] .
    \label{eq:dtc agp2}
\end{align}
In the third term, $\mu$ labels which leg hosts the $Z$ operator. In the last term, $i$ labels which link hosts the $X$ operator, while $\mu \in \{0,1,2,3,4,5\}$ labels which of the 6 remaining legs hosts the $Y$ operator.

We now optimize our second-order ansatz. Let's simplify notation by using $B_1 \equiv 80  h_x^3 + 32  h_x K^2$, $B_2 \equiv -48 h_x^3$, and $B_3 \equiv -32 h_x^2 K$. One can then show that 
\begin{align}
    G_K &= (-1 - 8 h_x (2 h_x  \alpha_1 + B_1 \alpha_2)) \sum_{v,\mu} \starZs + 4K(2 h_x  \alpha_1 + B_1 \alpha_2) \sum_i X_i \nonumber \\
    &+ (4 K (2 h_x  \alpha_1 + B_1 \alpha_2) - 12  h_x  B_3 \alpha_2) \sum_i \starXstarZs \nonumber \\
    &+ (4 h_x (2 h_x  \alpha_1 + B_1 \alpha_2) - 
    4 h_x  B_2 \alpha_2 - 4 K  B_3 \alpha_2) \sum_{v,\mu>\nu} \starYYZs \nonumber \\
    &-2 K  B_2  \alpha_2 \sum_{v,\mu} \starIXs + (2 K  B_2  \alpha_2 + 4  h_x  B_3  \alpha_2) \sum_{i,\mu>\nu} \starXstarYYZs \nonumber \\
    &+ 8  h_x  B_2  \alpha_2 \sum_v \starYs + 4  K  B_3  \alpha_2 \sum_{v,\mu>\nu} \starXstarXstarZs .
\end{align}
In the $-2 K  B_2  \alpha_2$ term, $\mu$ labels the leg with no $X$. In the next term, $\mu>\nu$ label the two legs with $Y$. In the final term, $\mu>\nu$ label which legs of the central vertex $v$ host the $X$'s connecting to the two outer vertices.
The action is then
\begin{align}
    \mathcal{S} &= \mathcal{D} N_v \Bigg[ (-1 - 8 h_x (2 h_x  \alpha_1 + B_1 \alpha_2))^2 + 2 (4K(2 h_x  \alpha_1 + B_1 \alpha_2))^2 + 2 (4 K (2 h_x  \alpha_1 + B_1 \alpha_2) - 12  h_x  B_3 \alpha_2)^2 \nonumber \\
    &+ 6 (4 h_x (2 h_x  \alpha_1 + B_1 \alpha_2) - 
    4 h_x  B_2 \alpha_2 - 4 K  B_3 \alpha_2)^2 + 4 (-2 K  B_2  \alpha_2)^2 + 30 (2 K  B_2  \alpha_2 + 4  h_x  B_3  \alpha_2)^2 \nonumber \\
    &+ (8  h_x  B_2  \alpha_2)^2 + 6 (4  K  B_3  \alpha_2)^2 \Bigg] .
\end{align}
Extremizing this action gives the protocol
\begin{align}
    \alpha_1 &= -\frac{120 h_x^4+451 h_x^2 K^2+64 K^4}{8 \left(192 h_x^6+1567
   h_x^4 K^2+662 h_x^2 K^4+64 K^6\right)} , \nonumber \\
   \alpha_2 &= \frac{3 h_x^2 + 4 K^2}{16 \left(192 h_x^6+1567 h_x^4 K^2+662
   h_x^2 K^4+64 K^6\right)} .
   \label{eq:dtc opt alpha2s hz=0}
\end{align}
As before, we can estimate the proper scaling factor $\lambda_f$ using
\begin{align}
    \lambda_f = \frac{\int_0^{\infty} \alpha_1 dK}{\int_0^{4h_x} \alpha_1 dK} \approx 1.304,
    \label{eq:dtc agp2 lambdaf}
\end{align}
such that our driving protocol is
\begin{align}
    H_{\mathrm{ACD}} = H_{\mathrm{DTC}} + \dot{K} \lambda_f \mathcal{A}_K^{(2)} .
\end{align}
Since implementing this ``hemidiabatic gauge potential'' in regular approximate CD driving is numerically costly, we instead use the optimized coefficients to implement the pulse sequence from above on a larger $L_y=4$ cylinder.

\subsection{Preparation pulse sequence}

\begin{figure}[t!]
	\centering

    \includegraphics[width=\linewidth]{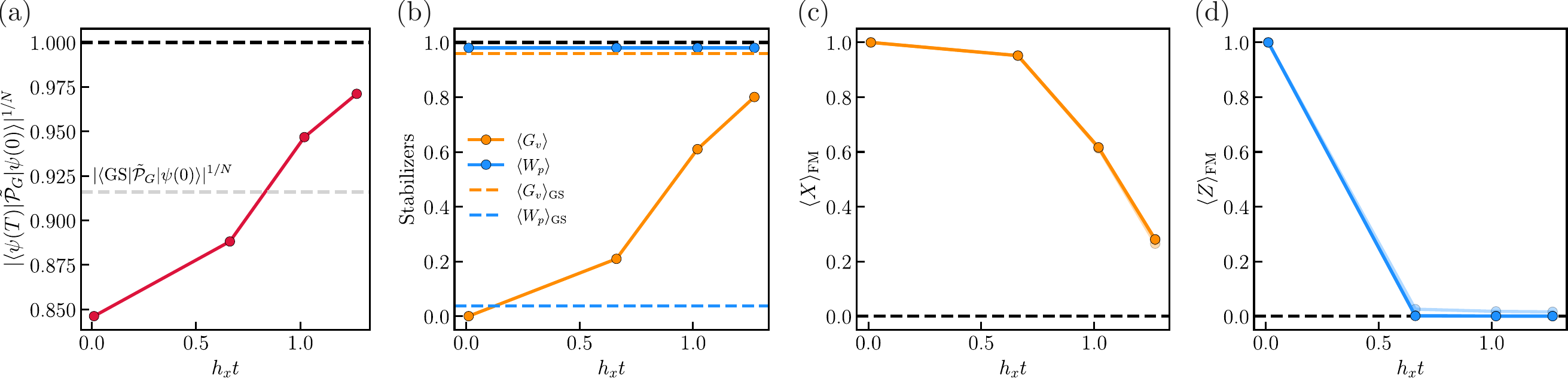}

	\caption{\textbf{Pulse Sequence in the DTC.} 
    We use an infinite cylinder with $L_y = 4$ and simulate a pulse sequence built out of 3 iterations of the cycle in Eq.~\eqref{eq:dtc cycle unitary}.
    We start from the product ground state with $K = 0$ and $h_z = 0.1 h_x$. 
    $x$ and $y$ are chosen for each cycle using the optimized driving parameters in Eq.~\eqref{eq:dtc opt alpha2s hz=0}.
    (a) We plot the fidelity per qubit as a function of time elapsed for each cycle. 
    The pulse sequence is able to prepare a quantum spin lakes state with about the same fidelity as a hemidiabatic sweep.
    (b) We also consider the stabilizer expectation values over the course of the pulse sequence.
    As expected, the Gauss law quickly increases while the Wilson loop remains exactly conserved.
    (c) The $X$ FM order parameter shows the $e$ anyons uncondensing during the pulse sequence in agreement with the Gauss law.
    On this larger geometry, we now measure the FM order parameter around three differently sized square loops, ranging from 1 enclosed plaquette in the lightest color (as in Fig.~\ref{fig:dtc acd driving}) to 9 enclosed plaquettes in the darkest color (see Fig.~4 in Ref.~\cite{Sahay_2023}).
    (d) The $Z$ FM order parameter begins at 1 due to the initial product state, but as soon as correlations develop the value shoots down to 0 as expected due to the lack of $m$ anyons.
    We also measure this order parameter around three differently sized square loops, ranging from 1 enclosed vertex in the lightest color to 9 enclosed vertices in the darkest color.
    }
	\label{fig:dtc pulse sequence}
\end{figure}

We now modify the pulse sequence from the text for use in the DTC. A single cycle of our protocol is given by the unitary
\begin{align}
    U_c = e^{-i x H_e} e^{-i y A_1} e^{2 i x H_e} e^{-i y A_1} e^{-i x H_e} ,
    \label{eq:dtc cycle unitary}
\end{align}
where 
\begin{align}
    A_1 = \sum_{v,\mu} \starYZs .
\end{align}
Rather than simply evolving under the starY Hamiltonian,
we now dress each time step with $H_e$ unitaries that generate the higher-order AGP terms via the Magnus expansion. By setting $h_z = 0$, we ensure that $W_p$ is always conserved and $m$ anyons remain exactly frozen. 

The parameter $x$ is defined by $x = \sqrt{-2\alpha_2/\alpha_1}$ where $\alpha_1,\alpha_2$ are the driving parameters found from optimizing the second-order AGP. With only this information, the protocol above is able to approximate the contributions of terms beyond second order. For the DTC, we consider constant $y$ and scan across various values to optimize the final fidelity, Gauss law, or other figure of merit.
Due to the computational intensity of these tensor network simulations, we leave numerical optimization to future work and instead show the increase in fidelity from the analytically optimized $\alpha_1,\alpha_2$.

In Fig.~\ref{fig:dtc pulse sequence}, we numerically evolve under such a drive protocol in the DTC. 
We take the sweep interval $K \in [0,4h_x]$ and divide it into 3 equal steps such that each cycle is assigned the $K$ value at the start of the respective interval. 
In this way, we can simulate the $K$ dependence of a quench sweep like in Fig.~\ref{fig:dtc acd driving}. 
The value of $x(K)$ for each cycle is thus determined using $\alpha_1(K),\alpha_2(K)$, and the optimal $y$ is found to be $-2.17 \times 10^{-2}$. 
All figures of merit and order parameters surpass the quench values of first order driving in Fig.~\ref{fig:dtc acd driving} and reach the spin lake values reached during hemidiabatic sweeps, again showing that our protocol does not require the high Floquet frequencies of the protocol from Ref.~\cite{Claeys_2019}.
As such, our protocol is able to include higher order terms in the AGP using just evolution under $A_1$ and $H_e$.

Although the state reached by this protocol is a quantum spin lakes state, comparing these results to those in Fig.~\ref{fig:dtc acd driving} and Ref.~\cite{Sahay_2023} reveals that no speedup in preparation is achieved. The long evolutions under $H_e$ set by $x(K)$ increase the protocol time to equal that of an undriven sweep. As such, using CD-inspired pulse sequences for a preparation speedup in the DTC requires further optimization of $x,y$, similar to our results for the Rydberg ruby lattice. We leave such optimizations to future work.

\end{document}